# European banks' business models and their credit risk: A cluster analysis in a high-dimensional context

Matteo Farnè [a], Angelos T. Vouldis [b, *]

[a] Department of Statistical Sciences, University of Bologna, Via Belle Arti 41, 40126 Bologna, Italy

[b] European Central Bank, Sonnemannstrasse 22, 60314 Frankfurt am Main, Germany

**Abstract**

In this paper, we investigate the credit risk in the loan portfolio of banks following different business models. We develop a data-driven methodology for identifying the business models of the 365 largest European banks that is suitable for very granular harmonised supervisory data. Our dataset allows us to take into account the full range of the activities in which banks are involved. The proposed method combines in an optimal way data clustering, dimensionality reduction and outlier detection. We identify four business models and exclude as 'outliers' banks that follow idiosyncratic business models. Furthermore, empirical evidence is provided that banks following different business models differ significantly with respect to the credit risk they undertake in their loan portfolios. Traditional commercial banks are characterized by the lowest levels of credit risk while the loan portfolios of securities holding banks are riskier compared to the other banks.

---

[*] Corresponding author.
*E-mail address:* Angelos.Vouldis@ecb.int (A. Vouldis).

# 1. Introduction

This paper applies a novel methodology to classify all systemically important Eurozone banks into business models, utilising a uniquely granular dataset. Our emphasis is on the post-financial-crisis banking system of Eurozone countries. We aim to overcome the limitations of the business model clustering approaches used in the literature that are based on narrow sets of broad pre-defined variables and do not fully capture the range of activities performed by banks. Our methodology combines optimally classification, dimensionality reduction, and outlier detection at the same time and accounts for the existence of idiosyncratic mixtures of banks' activities present in banks' business models (as noted e.g. in Mergaerts and Vennet, 2016). Furthermore, we contribute to the non-performing loans literature by providing empirical evidence that the business model choice can explain the differential credit risk of banks' loan portfolios in addition to the usual country- and bank-specific determinants.

The concept of 'business model' is increasingly used to refer to the heterogeneous mixture of activities to which banks engage and the difference in risk-return outcomes that this entails throughout different phases of the financial cycle (e.g. Yellen, 2012; Carney, 2015; Draghi, 2016). There is a burgeoning literature which proposes methods to classify banks into business models (Roengpitya et al., 2014; Ayadi et al., 2015; Köhler, 2015; Mergaerts and Vennet, 2016; Hryckiewicz and Kozlowski, 2017; Lucas et al., 2017). However existing empirical studies usually concentrate on a limited set of pre-selected dimensions when classifying banks into business models. In contrast, our method is data-driven and instead of relying on researcher's priors utilizes a unique, harmonised dataset detailing the activities undertaken by banks with an unprecedented level of granularity.

The paper contributes to filling the gap in the literature regarding banks' business models in the following ways. First, it formulates a methodology for identifying business models using granular data. Our proposed methodology combines optimally classification with dimensionality reduction allowing us to infer the factors which primarily determine banks' business models. It also incorporates an outlier detection component, allowing to identify banks which follow especially idiosyncratic business models, and whose inclusion into the normal clusters of banks would 'contaminate' the sub-sample and affect the results of analyses on differential risk or performance across business models.

Second, a unique data set, which has been made possible by the centralisation of supervision in a subset of countries within the European Union and the collection of supervisory data using harmonised definitions, is utilised. Our granular dataset comprising in total 1039 variables allows us to avoid biased classifications due to mismeasurement which could arise when broad categories are used, as will be explained below. In addition, the emphasis on the cross-sectional granular dimension of the input dataset is justified due to the slow-changing nature of banks' business models.[1]

Third, we identify the distinctive characteristics pertaining to each business model as regards the credit risk present in their loans portfolio, complementing the existing literature on the determinants of non-performing loans (Nkusu, 2011; Louzis et al., 2012; Klein, 2013; Ghosh, 2015; Anastasiou et al., 2016). According to our knowledge this is the first econometric study on the link between banks' business models and the credit risk of their loans portfolio. We find that there are statistically significant differences on the credit risk of banks following discrete business models, even after controlling for the potential endogeneity of the business model choice.

Our paper complements and leverages previous research on banks' business models. During the last couple of years a number of studies appeared that derive the business model classification from a narrow set of predefined variables (Roengpitya et al., 2014; Mergaerts and Vennet, 2016; Lucas et al., 2017), usually dictated by data availability. Other studies use a classification provided by the data provider (e.g. Köhler 2015, Becchetti et al., 2016) like the Bankscope's 'specialisation' attribute or focus on bank' ownership i.e. public or private (Eichengreen and Gupta, 2013). However, it has been already noted in the literature that the concept of business models, at least in Europe, belongs to a continuum (Mergaerts and Vennet 2016) if a restricted number of dimensions is used to classify banks.

Some specific examples will be used to illustrate the need to employ a granular dataset in order to distinguish banks into business models. For example, in previous studies broad categories like 'loans' or 'deposits' are used to identify the mix of activities into which banks engage. Therefore, a bank that holds an amount of loans to real economy agents above a certain threshold would most probably be classified as a traditional bank rather than an investment bank, however one may have to look also at off-balance sheet items, like loan commitments, to identify correctly a bank's involvement in financing the real economy. For example, BNP Paribas one of the largest bank in Europe, has off-balance sheet loans commitments and financial guarantees to the real economy representing 56% of the loans to the real economy which it holds in the balance sheet.[2] Therefore, the classification of such banks depends crucially whether one relies on broad aggregates (like the value of loans in the balance sheets) because these aggregates may not represent fully the range of the banks' activities, or whether one adopts a more granular view that permits a robust identification of the range of activities in which the bank is involved (e.g. by taking into account information on off-balance sheet financing).

---

[1] Our classification results remain unchanged to a level higher than 95% when a different reference date within the 2014Q4 – 2015Q3 time period is used. This is expected as in a short time frame the composition of banks' activities is not expected to change significantly. This is consistent also with the approach adopted in the literature. For example, Lucas et al. (2017) assume a fixed cluster assignment although their data set spans a larger time period while Mergaerts and Vennet (2016) find that 'between' variation (differences across banks) exceeds the 'within' variation (changes over time within banks) for a sample of banks from 30 European countries. Finally, also studies which define business models with respect to governance structures assume constancy of business models (e.g. Becchetti et al., 2016).

[2] Loans to the real economy equal EUR 712 bln while the sum of loan commitments and financial guarantees to the real

The measurement of banks' involvement in real sector lending is also highly affected by credit risk conditions. As a result, traditional banks with large volumes of non-performing loans could be mistakenly perceived as being less focused on providing loans to the real economy, because in the balance sheet statements the amount of net loans is given i.e. the amount of loans after deducting allowances. This is the variable used in the literature, however when banks increase allowances this amount will diminish as a percentage of assets. The statistics for the Greek banks illustrate clearly this point: between end-2009 and end-2016 their amounts of gross loans to the domestic real economy remained almost unchanged (from EUR 188 bln to EUR 185 bln), however net loans decreased by 20% (from EUR 177 bln to EUR 141 bln) due to much higher provisions.[3] As a result, one could think that they decreased their activities in the real economy during this period, however this results is an artefact of their higher provisions due to realised credit risk and not of a change in their business model.

In addition, information on the use of derivatives, which in existing studies is found to be a distinctive element of different business models, could easily be distortive, if one does not incorporate information on both carrying amounts and notional values, or uses information on only the asset or the liability side, or does not take into account the intended use of derivatives e.g. as hedging or trading instruments. If only one of these measures is taken into account, then a distorted view of the degree to which a bank uses derivatives will be obtained, affecting subsequently its classification into business model clusters.

Finally, classifications based e.g. on the class of a bank within a national financial system would probably also not be optimal. For example, the German 'Landesbanken' differ among themselves as regards the composition of their balance sheets e.g. their composition of funding sources. Consequently, lumping these banks into one category would distort results of an econometric analysis. For example, deposits from other banks or customers are generally an important source of

---

economy equal EUR 404 bln, according to the bank's 2016 financial statement: https://invest.bnpparibas.com/sites/default/files/documents/etats_financiers_31.12.16_en.pdf.

[3] Source: Bank of Greece statistics on the aggregate balance sheet of Greek credit institutions.

funding for Landesbanken, however the percentage varies considerably[4] and consideration of additional granular information on the remaining part of the liability side is needed in order to classify these banks meaningfully to a business model.

All the examples presented above show why using a narrow set of variables to classify banks into business models could be problematic and potentially misleading. Consequently, there is a high value added of using a granular data set as the multidimensional concept of a business model requires detailed information to be captured and an identification methodology which can handle such granular information. A business model identification method which uses granular input data needs to perform not only clustering, but also dimensionality reduction and the identification of factors, in addition to including outlier detection in the sense of identifying banks following very idiosyncratic business models. Besides granularity, other features of our dataset render it ideal for the purpose of identifying banks' business models. First, the dataset includes all large banks in Europe, with harmonised data consolidated at the prudential perimeter. This means that it provides the most reliable information on banks of systemic importance operating in Europe without exclusions which may bias the results e.g. inclusion of only listed banks. Furthermore, the prudential perimeter of consolidation as opposed to accounting consolidation or use of stand-alone data is optimal when considering banks' business models from a financial stability perspective because it provides simultaneously a consolidated view of banks activities while abstracting from non-banking activities which possess very different risk characteristics than banking. Finally, it is important to note that all banks in our sample operate under the same regulatory environment, based on the transposition of Basel III in Europe via the CRD IV package, therefore the effect of regulation on the composition of the portfolio of their activities is similar.

The paper is structured as follows. Section 2 presents a review of the literature on the concept of the business model and on relevant empirical studies on banking. Section 3 describes the input set and presents the clustering methodology. Section 4 presents the results and provides a discussion about the identified business models. Section 5 conducts an econometric analysis of the link between business models and credit risk in the loan portfolio. Finally Section 6 concludes.

---

[4] Specifically, some Landesbanken rely almost exclusively on deposits for their funding, either from other banks or from customers while for some other this is not the case e.g. the Oldenburgische Landesbank had according to its 2016 financial statement more than 85% of its funding via deposits, while this percentage is less than 50% for Landesbank Baden-Württemberg – again as shown in the bank's 2016 financial statement.

## 2. Review of the literature

The literature of business models in banking had been until recently driven mainly by the concept of 'strategic groups', introduced by Hunt (1972). The concomitant concept of 'mobility barriers' (Caves and Porter 1977) had been introduced to explain persistent performance differentials between firms within one industry and also applied in banking (Amel and Rhoades 1988; DeSarbo and Grewal 2008, Halaj and Zochowski 2009; Mehra 1996; Reger and Huff 1993; Tywoniak et al. 2007). Clustering methods are applied to identify strategic groups and consequently performance indicators are examined to assess whether performance differences exist. Data constraints are dictating the choice of the dimensions along which clustering is performed while the focus has been always in national or regional banking systems. Expert judgment is used extensively in the selection of the input set.

### 2.1. *Empirical analyses of banks' business models*

In the last years there is an expanding number of empirical studies of banking systems based on the business model concept. Roengpitya et al. (2014) (RTT henceforth) provide a clustering method to distinguish an international sample of banks according to their 'business models', based on the Ward's algorithm (1963), by using a selection of asset and liability variables (*choice* variables). Specifically, RTT define business models based on eight balance sheet ratios (loans, securities, trading book, interbank lending, customer deposits, wholesale debt, stable funding and interbank borrowing) which are interpreted as "reflecting strategic management choices" that leverage on the strengths of each organisation. They test their model on 1299 data points from 222 banks operating in 34 countries across the period 2005-2013, identifying three main business profiles: the Retail-funded, the Whole-funded and the Trading one. Finally, they provide a description of the bank performance for each business cluster by using a selection of key balance sheet ratios (*outcome* variables).

Ayadi and de Groen (2014) and Ayadi et al. (2015) also define banks' business models based on their activities. They examine a set of European banks (covering 80% of all banking assets in EEA in Ayadi and de Groen (2014)) and select a small number of dimensions (specifically, loans, trading assets, liabilities to other banks, customer deposits, debt liabilities and derivative exposures[5]) to perform hierarchical clustering. Both RTT and Ayadi et al. clearly distinguish between "activities", which determine the business model, and "outcomes", the latter measured by profitability and performance indicators.

---

[5] In Ayadi et al. (2015), liabilities to other banks and customer deposits were substituted by customer loans, because the expansion of the dataset compared to Ayadi and de Groen (2014) imposed more constraining data limitations.

Ayadi and de Groen (2014) find that retail banks, specifically the two business models which they label "diversified retail" and "focused retail", exhibit lower leverage compared to the business models of "investment" and "wholesale" banks which depart from the traditional intermediation function. However, this lower leverage is not reflected in their risk-adjusted ratios which are similar across business models, which can be interpreted as a symptom of "risk optimisation" on the part of sophisticated large banks. The results regarding performance are not clear-cut, also because of the volatility in the time dimension, however, it seems that "diversified retail" banks performed overall better than the other business models when taking into account both the pre- crisis and the crisis periods. Ayadi et al. (2015) expand the sample compared to Ayadi and de Groen (2014) and cover 95% of all banking assets in EEA. They also find that retail banks are less risky than wholesale and investment banks when using market measures of risk. RTT identify three business models and also find that their "retail-funded" banks perform better than "wholesale-funded" and "trading" banks while "trading" banks hold the higher levels of capital. Other studies which follow this line of research include Köhler (2015), Mergaerts and Vennet (2016), Hryckiewicz and Kozlowski (2017) and Lucas et al. (2017).

The important distinction between "choice" and "outcome" variables, aims to differentiate the set of variables reflecting strategic choices from the differential performance which is investigated ex post. The empirical strategies followed to classify banks into strategic groups usually focus on balance sheet "choice" variables (Amel and Rhoades 1988; DeSarbo and Grewal 2008; Mehra 1996). Halaj and Zochowski (2009) include additionally income and cost components, however this expansion of the type of variables is justified as a proxy for the unavailability of granular balance sheet breakdowns. Finally, Tywoniak et al. 2007 use also customer satisfaction ratings, although this seems to be better suited as a performance variable which could be investigated ex post.[6]

## 2.2.  *Identification methodologies*

The identification of banks' business models requires the use of clustering methods that are known to belong to the class of unsupervised learning methods (Hastie et al. 2009). Agglomerative hierarchical methods like the Ward's clustering method (Ward 1963), which minimises the variance within clusters, rely on expert judgment. These methods are not suitable in a high dimensional context as it is not so easy to characterize clusters based on a large number of variables. This method is employed by RTT who select a priori subsets of eight variables representing bank assets and liabilities, excluding highly correlated variables.[7]

---

[6] Reger and Huff (1993) should be considered separately in this strand of the literature as they focus on the cognitive dimension of the managers and utilises data originating from interviews with bankers. As regards, the criteria used to determine the differences among strategic groups, DeSarbo and Grewal (2008) include performance, efficiency and size in the outcome set. Halaj and Zochowski (2009) also incorporate risk indicators ('irregular loans') arguing that this allows to position banks in a risk-return space, an idea which is especially relevant for the banking sector.

In the direction of classifying large dimensional objects, clustering methods which incorporate a dimension reduction process have also been proposed. The dimensionality reduction component is critical for the set-up where the input set is granular and relatively large compared to the number of entities to be classified. The most obvious way by which dimension reduction issues can be incorporated into a clustering methodology could be through applying a principal component analysis (Hotelling 1933) or a classical factor analysis before conducting the clustering. Consequently, a standard unsupervised clustering algorithm like the Ward's one on the obtained principal components or factors can be applied. This approach is called *tandem analysis* (Arabie and Hubert 1994).

However, as pointed out in De Soete and Carrol (1994) and De Sarbo et al. (1990), this approach may not be the most efficient for classification. The dimensions identified by the principal components or the factor analysis are not necessarily the ones that maximise the distance among the latent clusters identified by the second step. Performing the dimensionality reduction in a separate, initial step may mask or obscure the true cluster structure of the data, since it classifies the objects according to directions which are not optimal for discriminatory purposes.

An effective solution which incorporates dimension reduction into the class of partitional clustering techniques labelled as "k-means" (MacQueen, 1967) is provided by Vichi and Kiers (2001), who develop the factorial k-means algorithm, where a subspace is defined such that the projected data points on this subspace are closest to the centroids. As the name of the procedure suggests, it involves both factor analysis (reducing dimensionality) and k-means procedure (clustering objects and finding out their centroids in this low-dimensional subspace). We adopt an enhanced version of this clustering approach which seems to optimally combine the two essential features, dimensionality reduction and clustering. We incorporate in the clustering algorithm an intrinsic procedure to identify outliers within clusters, using the factor scores obtained by the iterative algorithm.

Before proceeding to the detailed description of the methodology used, we mention the alternative family of methods based on finite mixture models, which has also been used in the literature to identify banks' business models (Lucas et al. 2017).[8] Density-based approaches are computationally heavy in large dimensions, since they are likely to result in a large number of clusters. In addition, distribution hypotheses on economic data are potentially more distortive in the banking context than, for instance, on genetic data with pre-defined labels (see Lin et al. 2016 and Murray et al., 2014a,b). Furthermore, due to the distribution assumptions made in finite mixtures models, the outlier detection as regards the banks' business models cannot readily incorporate information from all input dimensions. This may lead to the "contamination" of the identified banks' clusters with very idiosyncratic institutions potentially distorting the results. In addition, several low rank spaces are identified instead of one in cited works, with the exception of Murray et al. (2014b). This further complicates the description of identified outliers via those methods.

---

[7] The number of clusters is chosen using the pseudo F-index, as proposed in Calinski and Harabasz (1974), which quantifies the trade-off between parsimony and ability to discriminate between clusters.

Compared to the finite mixtures approach as used e.g. in Lucas et al. (2017), our enhanced clustering methodology offers the possibility to utilise a granular set of input dimensions, without the need to commit to a restricted set of inputs. The "trimmed" factorial k-means approach relies on a least squares algorithm which is effective in large datasets, because we identify only one latent space in place of several ones with a constrained distribution. In addition, a distribution-free approach lets the data speak with respect to the shape of clusters while also identifies banks lying far from the estimated clusters ("radial" outliers) without relying on parametric assumptions.

## 3. Methodology for identifying business models

### 3.1. Input set

We use a set of proprietary supervisory data which are collected in the context of the ECB Supervision. These data have been developed by the European Banking Authority (EBA) and employ harmonised definitions, thus representing an ideal set for a comparative analysis across countries. The availability of harmonised data across jurisdictions represents a necessary precondition when attempting to classify banks into respective business models.

We focus on Financial Reporting (FINREP) variables, providing a detailed decomposition of the balance sheet. FINREP is a standardised EU-wide framework for reporting accounting data, with a prudential scope of consolidation. Our sample consists of 365 banks residing in the 19 Eurozone countries. All systemically significant banks, as defined by the ECB Supervision (based on their absolute and within-country, relative size), are included in this sample. Our data set is cross sectional with reference date end- 2014. The variables included in the input set could be interpreted as the "choice" variables, reflecting banks' choices about the set of activities in which they are involved.

---

[8] The standard reference on finite mixture models is McLachlan and Peel (2000). Recently, the literature has provided robust versions estimating mixtures of multivariate skew-normal (Lin et al. 2016) and skew-t distributions (Murray et al., 2014a) by maximum likelihood and the EM algorithm respectively. A distribution-free alternative is provided in Yang et al., 2017 using trimmed likelihood. Lucas et al. (2017) estimate via EM dynamic mixtures of normal or t distributions with time-varying means and possibly covariance matrices and find that the choice of Student's t causes clusters to be more robust to outliers due to fat tails.

In particular, our input set contains information on the banks' balance sheet composition under four types of breakdowns, specifically accounting portfolios, instruments (loans, securities etc), counterparties (households, non-financial corporations etc) and products (mortgage loans, credit cards etc). Appendix A presents a detailed description of the input data set.

Each of the 1039 initial variables is standardised using total assets as the scaling factor, except from the 'total assets' variable which is normalised using its maximum value within the sample. Therefore, a 'size' variable is retained in the initial data set while almost all the remaining variables lie in the interval [0,1] since they are expressed as a percentage of size.[9] Standardisation is used because we define business models with respect to the composition of banks' activities, consistently with the literature reviewed above, and to avoid a dominance of the classification procedure by the large banks. Our results are invariant if the to-asset-ratios of any balance sheet variable and the relative size of any bank with respect to the maximum asset size in the sample are kept constant.

In this initial set there is a number of variables which are highly correlated and information which is redundant. Correlation is not per se an issue for the application of our clustering algorithm. However, given that we run the clustering algorithm with a number of different initialisations in order to search in the space of solutions and that we use the covariance matrix of the input set for the initialisation, the presence of nearly duplicated variables among the input data is not desirable.

Therefore, we follow a procedure to minimise the presence of very correlated variables in the input data set. The procedure consists of selecting the variables that should remain in the input set according to their 'importance', which is measured for each variable as the sum of the absolute correlations respect to all the others. This is a pre-processing step intended to avoid nearly duplicated variables by detecting pairs of variables that show a sample correlation very close to 1. Taking also into account the fact that some of the initial variables were very sparsely populated, we narrowed down the initial set of 1039 variables into a final input set of 382 variables (see Appendix B).

## 3.2. Clustering method

The statistical clustering problem can be defined as follows. Given a $n \times p$ data matrix $X$, where $n$ is the number of banks (objects or observations) and $p$ is the number of the variables, we would like to classify the banks into distinct clusters which contain objects which are 'close' in a statistical sense. Each cluster would represent a specific business model. The salient feature of our problem is that the dimension is relatively high compared to the number of objects: $p > n$. This feature is not common in similar classification problems; typically the objects which are to be classified are many more than the number of observed variables. Therefore our problem belongs to the field of clustering in high dimensions. In addition, the absolute number of dimensions necessitates the use of data reduction techniques in order to compress the large initial data set into meaningful composite variables.

---

[9] It should be noted that there exist few variables presenting values higher than unity, like notional amounts of derivatives.

As explained in Section 2.2, existing studies on clustering banks do not provide a readily available suggestion on how to approach the clustering problem in a high dimensional space. Specifically, it is not clear how strong is the impact of distribution assumptions for the data in our large-dimensional setting. Consequently, density-based clustering methods, which are based on normal or Student's t mixtures, may hinder the interpretation of results. In addition, there is no clear rationale for defining ex ante the distribution shapes of variables across business models. Alternative methods, like hierarchical and partitioning (i.e. centroid-based) methods, do not provide any dimension reduction by themselves.

Let us call $r$ the latent rank (i.e. the dimension of the reduced space) and $c$ the number of clusters. In formal terms, the model involves the minimization of a measure of the following matrix

$$\boldsymbol{XA} - \boldsymbol{U\bar{Y}} \tag{1}$$

where $\boldsymbol{A}$ is a p × r column-wise orthonormal matrix (coefficients matrix), $U$ is a n × c membership (or grouping) matrix such that $u_{ij} = 1$, if and only if $o_i \in Pj$, where $o_i, i = 1,..,n$, is the i-th observation and $Pj$, $j = 1, \ldots, c$ is the j-th cluster. The c × r matrix $\bar{Y}$ contains the centroids of the cluster in the low rank space. The left term of this expression represents the projections into the factor space of the original objects (the transformed variable space or low rank or reduced space), while the second term represents the centroids of the clusters.

Equation (1) lies in the low-dimensional space spanned by the columns of the column-wise orthonormal matrix $\boldsymbol{A}$. Consequently, the model can be specified as follows

$$\boldsymbol{XAA'} = \boldsymbol{U\bar{Y}A'} + \boldsymbol{E} \tag{2}$$

where $E$ is a residual matrix. Equation (2) describes the partition in the original space. The optimal partition therefore is sought by minimizing the function

$$\boldsymbol{F(A, U, \bar{Y})} = ||\boldsymbol{XAA'} - \boldsymbol{U\bar{Y}A'}||^2 = ||\boldsymbol{XA} - \boldsymbol{U\bar{Y}}||^2 \tag{3}$$

which can be equivalently expressed as

$$F(A, U) = || XA - U(U'U)^{-1}U'XA ||^2 \qquad (4)$$

since $\bar{Y} = (U'U)^{-1}U'XA$. This minimisation is performed under the constraints that $A'A=I_r$ and $U$ is binary with only one non-zero element per row. In geometrical terms, we seek for the orthogonal linear combinations of the variables (factors) which best partition the objects by minimising the least-squares criterion (Eq. 4) in this reduced space.

We follow a robust approach which belongs to the class of Alternated Least Squares (ALS) algorithms (Vichi and Kiers, 2001) which is explained in detail in Appendix C. A discrete clustering model and a continuous factorial model are specified *simultaneously* for our data set. So we perform at the same time data reduction (i.e. data synthesis) and variable selection by a single cluster analysis method, thus identifying the composite variables which most contribute to the classification of objects.

The strong consistency in statistical sense of the factorial k-means procedure is proved in Terada (2015). Underlying assumptions only require that the $p$-dimensional data vectors $X_i$, $i = 1, \ldots, n$, are IID with a common distribution $P$. The only constraint is that the random space spanned by the $p$ components of $X_i$ is not isomorphic to any random space of dimension $r$, where $r$ is the chosen latent rank. In high dimensions, it is clear that this is very unlikely to occur as $r$ is extremely small with respect to $p$. Therefore, consistency is ensured if $n \to \infty$. Note that the condition $p \geq n$ is not ruled out as long as the described constraint is satisfied.

The selection of the latent rank $r$ and the number of clusters $c$ is not straightforward. It has to be noted that these two parameters depend on each other. Specifically, the number of components $r$ cannot be larger than $c - 1$. The reason is that $Rank((U'U)^{-1}U'XA)) = \min(c - 1, r)$, therefore describing the low-dimensional space of the clusters using more dimensions than necessary does not seem to make sense. The process for selecting these parameters will be described in Section 4.1 since it combines statistical criteria and the aim of obtaining interpretable results.

### 3.3. *Robustified clustering with simultaneous radial outlier detection*

It is known a priori that some institutions in our dataset follow unique business models, e.g. functioning as central clearing counterparties, focusing exclusively on refinancing public sector loans etc. Therefore, there is a clear rationale for excluding these outliers from the clusters to avoid distortions of the final results. We want to make sure that the presence of such cases does not distort the classification of banks. Therefore we present a robustified version of the Vichi and Kiers (2001) procedure, which identifies clusters taking iteratively into account the presence of radial outliers. Our method is specifically intended to identify the so-called radial outliers, that is, observations deviating so much from assigned centroids to be considered external to assigned clusters. Therefore, our approach is robust both in the sense of measuring in a robust way the composition of the banks' activities, due to the granularity of the used dataset, and in avoiding the 'contamination' of the identified clusters with banks following particularly idiosyncratic business models.

As well established in literature (see e.g. Rousseeuw and Leroy, 2003), the most used method for detecting multivariate outliers is via the Mahalanobis distance, $D = \sqrt{(x-\overline{x})'S^{-1}(x-\overline{x})}$, with $D^2$ being asymptotically a chi-squared with $p$ degrees of freedom under the assumption of normality for $x$ ($S$ is the unbiased sample covariance matrix). Under the normality assumption, Hotelling (1933) showed that $t^2 = n(x-\overline{x})'S^{-1}(x-\overline{x})$, called Hotelling's T-squared, is proportional to $F_{p,n-p}$, where F is the Fisher's F. However, it is easy to see in our context this approximation cannot be used, because normality is not respected and the degrees of freedom $n-p$ would be negative, since $p > n$.

The trimmed k-means algorithm proposed in Cuesta-Albatos et al. (1997) could be used to detect anomalous data simultaneously with clustering. However, this method may be computationally intractable when both $p$ and $n$ are large while it does not offer a clear interpretation and visualisation of the identified clusters in large dimensions. In contrast, we would like to identify both partitions and outliers in a reduced space rather than in a p-dimensional space.

For this reason, a method is developed here to find the partition of the $100 \times (1-\alpha)\%$ most concentrated objects with respect to the scores in the low-dimensional space. Specifically, in the absence of any distribution assumptions we compare Mahalanobis distances across banks in order to exclude radial outliers i.e. banks clearly different from the rest. Mahalanobis distances are based upon $C_F$, the unbiased covariance matrix of factor scores estimated over the entire sample, because the heuristics based on Equation (4) do not explicitly address the possibility of significantly different covariance matrices across cluster.

In more detail, our problem may be stated as follows. Our task is to minimise $F(A, U, \overline{Y}) = ||XA - U\overline{Y}||^2$ under the constraints $\sum_{i=1}^{n}\sum_{j=1}^{c} u_{ij} = [(1-\alpha)n]$, $\sum_{j=1}^{c} U_{ij} \leq 1$ for each $i = 1, ..., n$, where the trimming proportion is $\alpha \in [0, 0.5]$. This trimmed problem may be practically solved under the framework of Rousseuw and Van Driessen (2000). As a subset selection step (H-step), we set 100 initialisers and compute the initial estimates of loadings, centroids and cluster memberships by the Alternated Least Squares algorithm of Vichi and Kiers (2001). As a concentration step (C-step), we compute the Mahalanobis distance at each observation and we exclude the $[\alpha n]$ observations with

the largest ones, since they are the observations that contribute the most to $\mathbf{F}(\mathbf{A}, \mathbf{U}, \overline{\mathbf{Y}})$, which is the variance within clusters. In this way, it is ensured that $\mathbf{F}(\mathbf{A}, \mathbf{U}, \overline{\mathbf{Y}})$ is decreasing at each step, such that the overall optimum is found out over the entire range of initializers which approximate the parameter space. The details of the outlier detection part of the algorithm are also described in Appendix C.

This robustified version of factorial k-means algorithm has two major advantages. First, the clusters are optimally shaped, given that distortions arising from the dimensionality reduction stage are avoided. Second, the outliers are automatically identified during the procedure by the clustering algorithm, with no need of applying any subsequent procedure.

## 4. Business models and their characteristics

### 4.1. Clusters and factors identified

The selection of the number of clusters and factors follows the methodology described in Appendix C. Figure 1 presents the Hartigan's statistic for different number of clusters and for two different strategies of selecting the number of factors which represent the lower and upper bounds, respectively. The first strategy is to keep the number of factors fixed and equal to two (2). The second is to use the maximum number of factors, $r = c - 1$. In addition, in both cases we plot the results both for the case where outlier detection is performed and for the case that no outliers are excluded. A common feature of all these lines is that Hartigan's condition is satisfied for $c = 4$ and therefore, we select 4 clusters.

**Figure 1:** Hartigan's statistic for different number of clusters and factors.

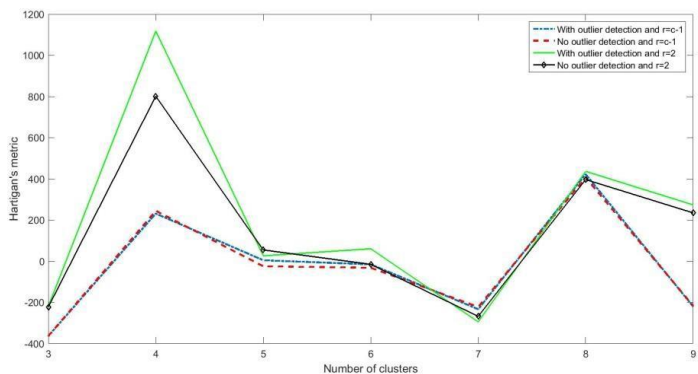

The maximum number of factors that could be used is $c - 1 = 3$, however when examining the singular values of the n x r matrix $(\mathbf{U}'\mathbf{U})^{-1}\mathbf{U}'\mathbf{X}\mathbf{A})$ (see Vichi and Kiers 2001), it is clear that clusters nearly fall in a subspace with dimensionality lower than 3.[10] Therefore we select r = 2, a decision which is further reinforced by our aim to obtain interpretable results.

In addition, we set α = 0.1, i.e. the 10% of the banks which are identified as outliers represent a separate group. The selection of the quantile value was chosen based on the examination of the set of banks which were selected in the outlier set and on the visual examination of results in the low dimensional space. Specifically, the chosen parameter values lead to a set of outlier banks that contains institutions with idiosyncratic features which are also distinctively far from the clusters' centroids in the low dimensional space, as it will be elaborated later. On the other hand, our classification results are not sensitive to this assumption in the sense that the cluster membership of all the remaining banks is not affected.

The two factors produced by the factorial k-means model consist of a 'level' factor and a 'contrast' (slope) factor. Intuitively, the first factor represents a measure of the presence of standard elements in banks' balance sheets, like loans, deposits, derivatives and issued debts, excluding trading assets. The second factor represents the contrast between loans and 'standard liabilities' (which include deposits and issued debt) therefore discriminates banks with respect to the imbalance of these standard items on their two sides of their balance sheet.

### 4.2. The business models identified

Figures 2 and 3 present the "median" balance sheet composition of banks in each cluster allowing a better understanding of the composition of activities which characterise the identified business models. To provide intuition we name the business models as follows:

*1.    Wholesale funded banks* are generally large banks, their asset side consists mostly of loans (second only to traditional commercial banks, see below), they rely much more than other types of banks on debt for their funding and less on household deposits (see Figure 3). These banks are characterised by far the largest use of derivatives, both for hedging and trading. This cluster contains the lowest number of banks: in total, 58 banks belong to this category.

*2.    Securities holding banks* hold a relatively large securities portfolio and cash buffer, fund themselves with deposits and do not use derivatives much. This business model holds the higher amount of cash, mainly to be able to carry out its trading activities.  This business model grants the lowest amount of loans (see Figure 2). The liability side of the securities holding banks looks pretty

---

[10] Specifically the singular values for three factors were 17.3, 3.7 and 0.4 while for the two factor case they were equal to 16.0 and 3.

'traditional' with a significant amount of deposits.[11] They are usually small, but this cluster is the most heterogeneous one as regards their size. The number of banks which follow this model is 86.

*3.   Traditional commercial banks* are medium-sized, have loans on their asset side more from all other banks (see Figure 2). These banks fund themselves more with deposits compared to all other business models (see Figure 3) and use derivatives primarily for hedging. They represent the textbook prototype of banks as financial intermediaries. The number of banks contained in this cluster is 77.

*4.   Complex commercial banks* are medium sized, possess a significant percentage of loans on their asset side but lower compared to traditional commercial banks because they also own securities to a larger extent, fund themselves mostly with deposits (but less than traditional commercial banks) and use derivatives mostly for trading purposes. This is a hybrid category, between traditional commercial and wholesale funded banks. It is the largest cluster and includes 108 banks.[12]

The numbers of banks that are classified in the various categories can be compared with those of RTT and Ayadi et al. (2015). RTT classify most out of the 67 European banks contained in their sample as "retail" with "wholesale funded" following and with "trading" banks representing the lowest number. Our results are in accordance with those of RTT when it comes to retail banks representing the majority of the banking population. There is however a discrepancy with respect to the relative numbers of "wholesale funded" and "securities holding" ("trading") banks, given that in our case the securities holding banks are more than the wholesale funded. This result could be driven by our extra category, namely the complex commercial banks which may include some banks that in RTT could have been labelled as wholesale funded. It may also be due to our larger sample that contains smaller banks that follow the "securities holding" business model. Our results are consistent with those of Ayadi et al. (2015) where retail banks are the majority followed by "investment" banks and with "wholesale" banks representing the minority. Given that the sample of Ayadi et al. is the largest from those compared here (with 2,542 banks from the EEA and Switzerland), it seems plausible that our results are somewhere between those of RTT who consider a small sample and Ayadi et al. in the sense that the importance of securities holding banks seems to increase as the sample gradually becomes larger. There is also a clear correspondence of our business models to those identified by Hryckiewicz and Kozlowski (2017).

---

[11] Therefore, it is clear that the two types which sometimes are lumped together as 'investment banks', namely the securities holding banks and the wholesale funded ones should be distinguished because their activities differ substantially.

[12] We have preferred to label this business model as "complex" rather than "diversified", as the latter label would imply that they are safer against risks. On the other hand, both names refer to the variety of the activities in which these banks are engaged to. The characterisation "universal" could also be fitting for this class of banks.

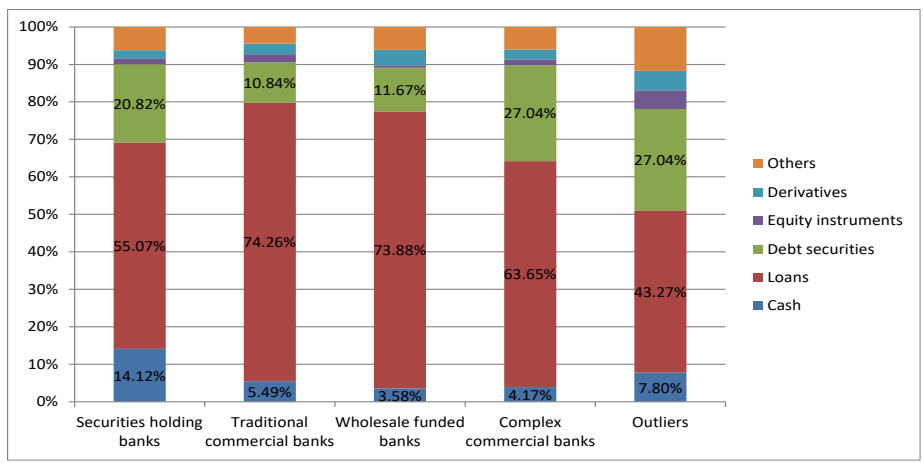

**Figure 2.** "Median" asset composition per business model.

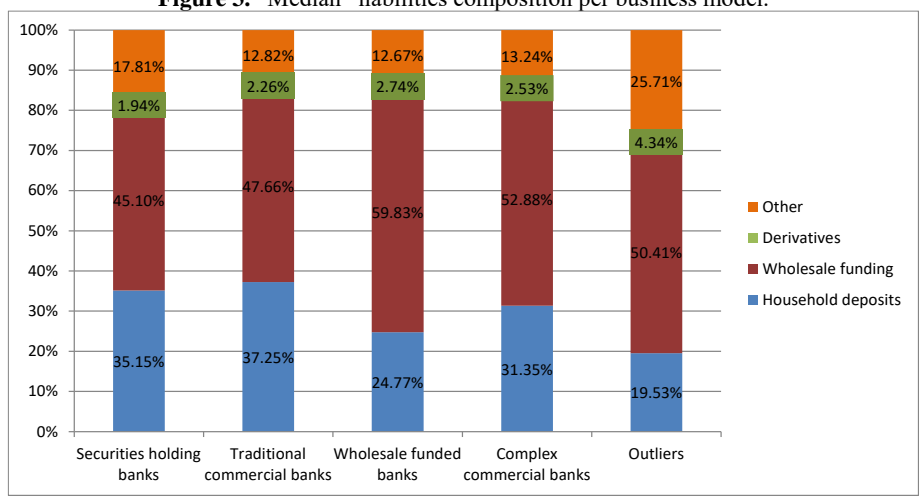

**Figure 3.** "Median" liabilities composition per business model.

Figure 4 presents a graphical illustration of the positions of all banks and clusters in the factor space while Figure 5 also includes the outlier category in order to show the position of outliers compared to the other classified banks. In addition, we report the centroids of the various clusters in the factor space below the Table.

Looking at the relative positions in the x-axis (level factor), the wholesale funded banks are clearly located leftwards compared to all other categories. Both types of commercial banks occupy approximately the same range across the x-axis while the securities holding banks are located on the right of all other types. This relative positioning conforms to the composition of the level factor as explained above. In particular, while both commercial banks and wholesale funded contain similar amounts of loans, deposits and issued debt, they differ with respect to the use of derivatives (higher for wholesale funded banks)[13] and this places the latter at the left end. On the other hand, the securities holding banks are at the right end of the x-axis given the large presence of trading assets and the relatively low presence of loans which lead to low absolute values for the Level factor.

**Figure 4.** Location of banks and clusters in the two-dimensional factor space. In the table below the graph, the coordinates of the centroids position in the factor space is presented.

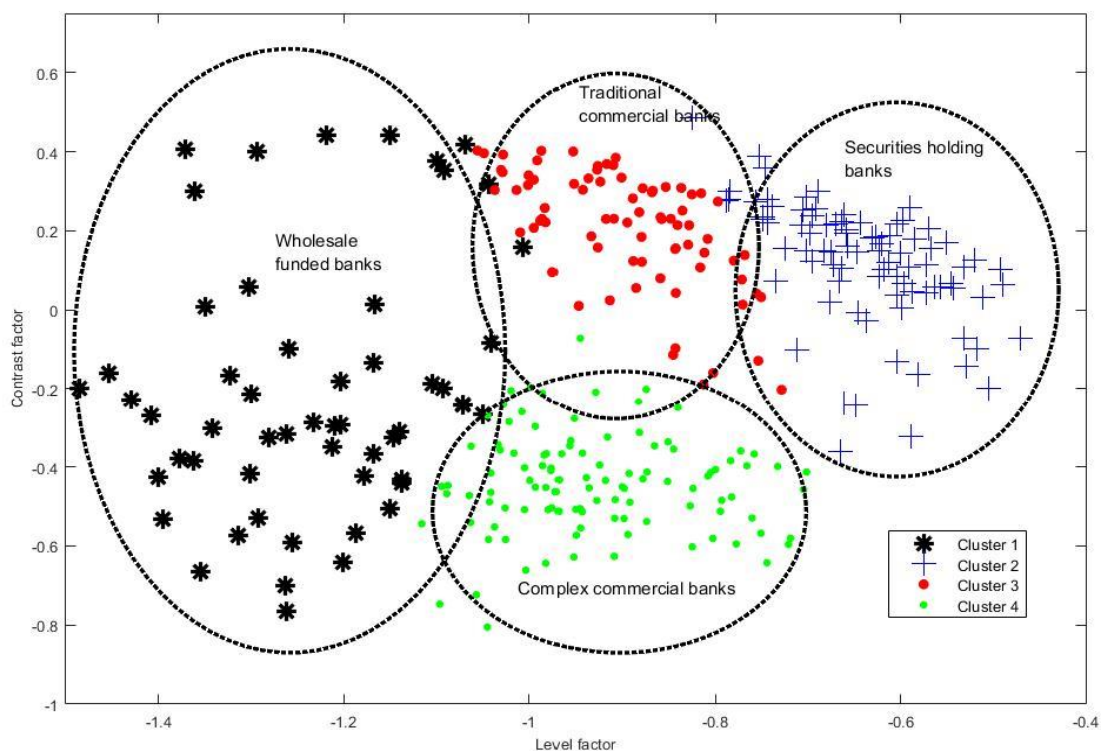

Centroids of clusters in the factor space

|  | Level factor | Contrast factor |
|---|---|---|
| 1. Whol. funded | -1.07 | -0.13 |
| 2. Sec. holding | -0.64 | -0.04 |
| 3. Trad. comm. | -0.82 | -0.08 |
| 4. Complex comm. | -0.96 | -0.16 |

With regard to the relative positions in the y-axis (contrast factor), it is interesting to note that the traditional commercial banks are located higher than the complex commercial ones, due to the more pronounced presence of loans on their asset side (loans enter with a positive sign in the contrast factor). On the other hand, wholesale funded banks occupy a wide range of positions with respect to the y-axis, reflecting varying contrasts between loans and standard liabilities. A seemingly counter-intuitive observation is the high contrast values exhibited by the securities holding banks, despite their relatively small percentage of loans (due to higher levels of trading assets and cash). This is explained when one considers their liability side which also includes a lower level of 'standard' liabilities compared to other types of banks. Specifically, the category "other liabilities" (besides deposits, debt and derivatives) is noticeably higher for securities holding banks. The trading activities are reflected in the high percentage values of this item comprising e.g. amounts payable in respect of future settlements of transactions in securities or foreign exchange transactions.[14]

---

[13] Specifically, the median carrying amount of hedge accounting derivatives on the asset side of the "wholesale-funded" cluster is 1.05% of the total assets while this number is less than 0.25% for the remaining clusters. The carrying amount of derivatives in the "Held for trading" portfolio is 0.96% for the "wholesale-funded" cluster while it is less than 0.57% for the remaining clusters.

Figure 5 provides a validation for the outlier component by showing the position of the outliers in the two-factor space. It is clear that the large majority of detected outliers are located distinctively apart from the other classified banks. Further insight into the composition of the outlier set can be gained by examining the initial classification of the banks which end up in the outlier set, before the outlier detection algorithm is applied. Specifically, the outlier banks set is composed of 18 banks which were initially characterised as *securities holding banks*, 13 banks which were initially characterised as *wholesale funded banks*, and 5 banks which were initially characterised as *complex commercial banks*. No bank from the *traditional commercial banks* category was reallocated as an outlier. Therefore, consistently with the qualitative observations above, mainly banks which depart from the standard model of a commercial bank were reclassified into the outlier category.

A closer examination of the set of 'outlier' banks reveals that it includes primarily small investment banks and specialised lenders. For example, we find in this set some local government funding agencies, specialising in providing financing for (semi-)publicly owned organizations and institutions refinancing loans to local public sector entities. Also included are some specialised subsidiaries of larger groups, a bank in a run-down mode and central clearing counterparties. Overall, included in this group are banks following clearly idiosyncratic business models.

---

[14] Fair valued financial commitments and guarantees are also included under this item – according to anecdotal evidence, such "other liabilities" is relatively more important for the other categories of banks, however this further decomposition is not readily available.

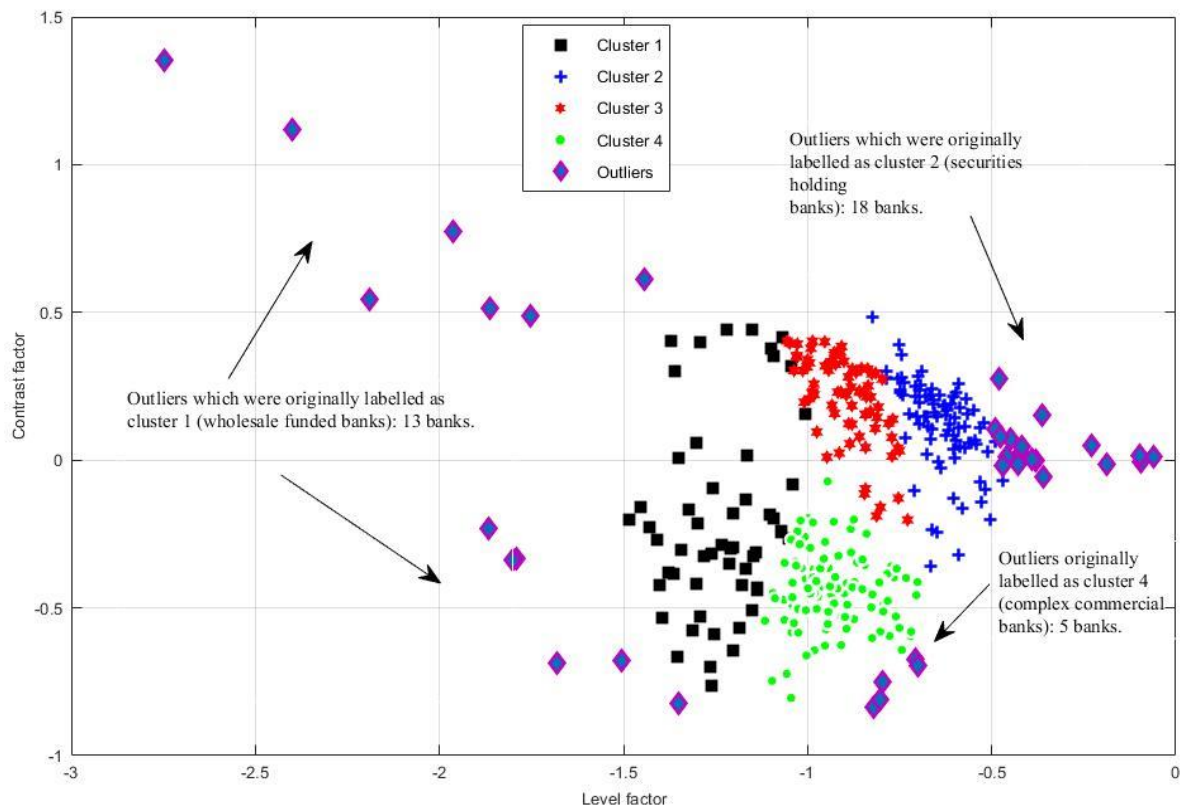

**Figure 5.** Location of banks and clusters in the two-dimensional factor space, including outliers.

## 5. Business models and credit risk

In this section, we investigate the credit risk present in the loan portfolios of the previously identified business models. We focus on two measures of ex post credit risk and condition on country-specific and bank-specific determinants aiming to investigate the additional explanatory power of the business model choice. The first measure of credit risk used is the default rate which is almost identical to the non-performing loan rate as used in the aforementioned studies. The default rate is a prudential measure which is defined in a harmonised way across eurozone banks based on the pan-European Capital Requirements Regulation (commonly known as the CRR).[15]

Furthermore and to enhance the robustness of our analysis, we use the alternative measure of net realised credit risk, which is the ratio of impaired and past due loans to total loans, but subtracting the provisions already accumulated by the bank both in the numerator and the denominator. In this sense, the second metric possesses also a forward looking component as it represents credit risk which has not been covered by provisions and may hit banks' capital directly. Our test hypothesis is that the business model choice is linked to the credit risk of the bank.

---

[15] We have preferred to use the default rate compared to non-performing rate as the definition has been harmonised across European countries for longer, meaning that harmonisation across countries could be more aligned. However, our results remain qualitatively similar when the non-performing loans ratio is used. This set of results is available upon request.

*H1(i): Banks following business model i exhibit differential credit risk compared to the banks that do not follow business model i.*

This formulation nests four hypotheses, one for each of the identified business models. At this point, we remain agnostic whether each model $i$, with $i = 1, ..., 4$ corresponding to the four identified business models, exhibits higher or lower performance. Instead we focus on investigating whether there is evidence simply for differential performance, a hypothesis which can be explained via the mobility barriers concept of the strategic groups literature (e.g. as in Portes 1979 and the literature which followed). Given our aim to test for the existence of differential performance characteristics, we run four separate regressions, one for each business model, and include each time a dummy variable for a specific business model.

### 5.1. OLS and fixed effects models

We perform two baseline estimations: one with country fixed effects and country-level clustering of the error terms and the second with OLS and country-level clustering of residuals. The results of these baseline estimations are subjected below to additional robustness checks. Specifically, the model estimated with OLS and assuming country level clustering of residuals is the following:

$$y_{j,t} = a_0 + a_1 Z_{k(j),t-1} + \beta X_{j,t-1} + cIND(i)_{j,t-1} + e_{j,t} \tag{5}$$

where the index $j$ spans the different banks in our sample ($j = 1, ..., 365$), $k$ spans the countries in the sample ($k = 1, ..., 19$), $k(j)$ is a function which maps each bank $j$ to the country in which it is operating, $y_j$ represents the dependent variable measuring credit risk, $Z_k$ is a vector of country-level variables for the country in which bank $j$ is operating, $X_j$ is a vector of conditional bank-specific variables, $IND(i)$ is an indicator variable that shows whether bank $j$ has been classified as following the business model $i$[16], $a_1$ and $\beta$ are vectors of coefficients and $c$ a coefficient. Time $t$ refers to end-2015 data and $t$-$1$ to end-2014 data. Finally, $e_{j,t}$ represents the residual term. The clustering of residuals across countries means that we assume

$$E[e_{j,t} e_{j^*,t} | X_{j,t-1}, X_{j^*,t-1}] = 0, \text{ except if } k(j) = k(j^*).$$

---

[16] One dummy is included each time the model is estimated, therefore the coefficient for each business model quantifies the average statistical difference of the effect for the banks following this specific business model when compared to the banks following the three remaining ones.

The coefficient of interest $c$, is the estimated effect of the business model $i$ on bank's credit risk. For example, a positive coefficient $\hat{c}$ means that the banks following the business model $i$ present on average higher credit risk compared to all the other banks.

The corresponding model with country fixed effects and country-level clustering of the error terms includes a separate intercept for each country and can be written as

$$y_{j,t} = a_{0,k(j)} + \beta X_{j,t-1} + c IND(i)_{j,t-1} + e_{j,t} \tag{6}$$

The same condition as in the case of the model estimated with OLS applies for the residual term $e_{j,t}$.

We follow the expanding literature on banks' credit risk determinants that utilises both macroeconomic and bank-specific variables. Our list of bank-specific variables is consistent with the specifications followed e.g. in Berger and DeYoung (1997), Louzis et al. (2012) and Nkusu (2011) where the determinants of non-performing loans in advanced economies are investigated. In addition, our specification is broadly consistent with the bank diversification and profitability strands of the literature (Albertazzi and Gambacorta 2009; Demirgüç-Kunt and Huizinga 1999; Elsas et al., 2010; Laeven and Levine 2007) which is related to our investigation. The bank traits which serve as conditional variables include the size of the bank, the ratio of deposits to liabilities to control for the funding structure of the bank, the leverage ratio (equity-to-assets), past growth of assets and income and its market share within its country. In order to avoid that outliers drive our results, we winsorise all used variables at the 5% level.

Some bank-specific variables that we use are also inputs to the classification of banks in business models. Therefore, we examine the multicollinearity of our baseline model variables by checking the VIF values and in all cases they are less than 1.1 meaning that multicollinearity can be safely ignored.[17] The absence of significant multicollinearity effects can be attributed to the multi-dimensional nature of our business model classification methodology. However, multicollinearity would represent an issue for the business model classifications that are based on a small number of pre-selected dimensions.[18] Therefore, due to our granular approach to business model classification we are able to control for such bank-specific variables which may differ within business models and whose omission may have led to omitted variable biases.

The first set of conditional variables takes into account scale effects which may impact the dependent credit risk variables. This set includes bank's size (following e.g., Beltratti and Stulz, 2012; Laeven and Levine, 2007), which could affect risk through too-big-to-fail (henceforth TBTF) effects (e.g. as in Boyd and Gertler, 1994; Ennis and Malek, 2005; Freixas and Rochet, 2013) or the existence of (dis)economies of scale. Additionally, we include variables measuring banks' growth of assets and operating income (e.g. as in Louzis et al., 2012; Ghosh, 2015).[19] Therefore, we condition both on cross-sectional and dynamic scale variables (Elsas et al., 2010). In this way one can discriminate between indirect size effects and direct effects due to expansion ('foot-in- the-door' strategy).

---

[17] These results are available upon request.
[18] For example, Hryckiewicz and Kozlowski (2017) do not include in their regression variables that are used for the clustering of banks.

Furthermore, we control for bank's funding composition and include the ratio of deposits to assets. Köhler (2015) does not find a statistically significant effect from non- deposit funding to banks' risk, however we would like to account for a variable which is typically used as a defining dimension in the existing business model literature in order to show that business model choice as defined here provides explanatory power over simple dimensions used to distinguish banks' business models. We also control for the level of capitalisation which reflects bank's risk for unexpected losses (Berger and DeYoung, 1997; Wheelock and Wilson, 2012). A weakly-capitalized bank would have stronger incentives to engage in higher risk activities that promise higher return.

Finally, it has been argued that the competition that each bank faces within its jurisdiction may have effects on its governance and efficiency (as for example in Berger and Mester, 2003) and subsequently on the risk of its portfolio. Therefore, we include the market share that each bank represents with respect to the total consolidated assets of the other banks that operate in the same jurisdiction as an explanatory variable.[20]

In all cases we use observations of the outcome variables for end-2015 i.e. one year after the reference date of the dataset used to identify the business models. In this way, potential endogeneity issues with respect both to the definition of the business model and the bank-specific determinants can be addressed. We also conduct a number of robustness checks of our results below, also utilising instrumental variables estimation to address potential endogeneity biases with respect to the business model variable.

Specifically, in the baseline model we mitigate for the potential endogeneity of the bank-specific variables by using year-long time lags. This also applies to the business model variable, which is based on end-2014 data i.e. precedes the outcome variables by one year. The theoretical justification which underlies our choice of variables aims also to address endogeneity caused by omitted variables.

---

[19] Due to data limitations we cannot go further back in time beyond yearly growth. However, we expect that the long-run bank's expansion would be correlated with the long-run changes in macroeconomic conditions for which we have conditioned already for a multi-year horizon, taking into account the developments during the crisis.
[20] As reported in the Consolidated Banking Statistics of the ECB:
https://www.ecb.europa.eu/stats/supervisory_prudential_statistics/consolidated_banking_data/html/index. en.html .

In the baseline regressions we treat the lagged business model variable as exogenous, which can be justified by the argument that one would expect that the business model is a feature which is constant in the medium term e.g. in the course of one year, and therefore there is no reverse causation stemming from the outcome variable. However, migration of banks across business models cannot be excluded and consequently concerns of endogeneity. Therefore, we investigate below the robustness of our results to endogeneity concerns regarding the business model classification.

In the specification estimated with OLS we include country specific variables such as GDP and unemployment measures in order to capture the effects of the macroeconomic environment in different time horizons.[21] Concretely, macroeconomic conditions are taken into account by including the y-o-y real GDP growth and the 2009-2015 change in unemployment, so as to capture both short-run macroeconomic effects and the long-run effects of the crisis.[22] When country fixed effects are used, additional country level effects can be captured e.g. the impact of national macro-prudential measures that could affect banks' risk (Altunbas et al., 2018).

*5.2.   Controlling for endogeneity*

We conduct two additional robustness checks based on alternative estimation methods.[23] The first aims to address endogeneity concerns with respect to the business model indicator. So far the assumption has been made that business models affect outcome variables but not vice versa while, in addition, we assume that there are no omitted characteristics which determine both the performance variables and banks' business models. The lag structure does not necessarily address such potential biases, given that the business model is a semi-stable characteristic for each bank, which may evolve throughout time even though only gradually. In addition, there could be unobserved features at the bank level, e.g. the quality of governance of an institution, that affect credit risk. This potential problem of endogeneity could be solved by using an instrument for bank's business models that would be a predictor of bank's business model but without having a direct effect on credit risk. Such instrument would proxy for the unobserved institutional features of a country that would affect the banks' choice of a business model e.g. with respect to the role of a country as a financial hub, past regulatory interventions[24], propensity of households to take mortgage loans[25], and would affect credit risk only via this channel and not directly (see Angrist and Pischke 2009).

---

[21] All country specific variables which are used, including those for the robustness checks, are explained in more detail in Appendix E.
[22] The results are qualitatively similar when we interchange the use of long-run and short-run measures for GDP and unemployment or when inflation is also included.
[23] An additional robustness check was to include efficiency as an explanatory variable, consistently with the 'bad management' and 'skimping' hypotheses of Berger and DeYoung (1997). The results are qualitatively similar while the statistical significance of the business model variable remains. These results are available upon request.

We proxy these unknown institutional features that determine banks' business model at the country level by the share of each business model within the subsample of banks of that country.[26] This instrumental variable is highly correlated to each bank's business model indicator however it can be plausibly argued that it is not a direct determinant of the credit risk to which each bank is exposed as it proxies for country level institutional factors while credit risk is mainly determined by macroeconomic and bank-specific determinants (see Louzis et al. 2012; Klein 2013). It can be also plausibly assumed that the share of banks following a specific business model in a country should be unrelated to unobserved features at the bank level e.g. the quality of governance of an institution. Consequently, the only reason for the relationship between the share of banks in a country that follow a specific business model and a bank's credit risk is the relationship between the share of banks variable and the bank's business model variable. Specifically, let $SHARE(i)_{k,t-1}$ denote the percentage of banks following business model $i$ in country $k$ in end-2014. The exclusion restriction is that $Cov(SHARE(i)_{k,t-1}, e_{j,t}) = 0$ while the data show also that the variable $SHARE(i)_{k,t-1}$ is a strong predictor of $IND(i)_{t-1}$[27], therefore it represents a valid instrument in our setting.

In addition, we also perform, as an additional robustness check, a backward stepwise regression approach, a methodological choice followed also in Aebi et al. 2012.[28] Specifically, we use backward stepwise regression with 20% significance level for removal from and 10% for addition to the model, with clustering of residuals at the country level.

### 5.3. Results

The regression results are presented in Tables 1 and 2 for the OLS and country fixed effects specifications, for realised credit risk and default rate, respectively. Furthermore, in Tables 3 and 4 the results of the robustness checks using instrumental variables and stepwise regression are shown for realised credit risk and default rate, respectively. As expected, the unemployment rate is a statistically significant driver of credit risk, consistently with Nkusu (2011), Louzis et al. (2012) and Anastasiou et al. (2016) among others. In addition, as regards the bank-specific variables, there is evidence that asset and income growth are negatively associated with credit risk, meaning that banks which are able to expand exhibit also lower levels of credit risk. This is in contrast to other studies which use data spanning a period including the peak of the most recent financial crisis (e.g. Klein 2013) and may be related to the capacity of the banks which were able to withstand the most intense period of the crisis to extend credit while being also exposed to lower credit risk. This is similar to results which utilise data only from the pre-crisis period e.g. Quagliariello (2007).

---

[24] It must be mentioned that our sample date (end-2014) corresponds to the initiation of the Single Supervisory Mechanism which encompasses all countries in our sample, so all banks are under the same regulatory regime. However still there could be regulatory interventions of the past which may have shaped banks' selection of business models.
[25] This analysis of potential endogeneity is at this stage simply a robustness check as there is no available literature on the determinants of banks' business models.
[26] We refer here to the institutional factors influencing the largest banks' business model in each country, as our sample comprises of the largest institutions.
[27] We have performed the F-statistics for the instrumental variables, which show that our chosen instrument is quite strong (in the sense of being correlated with the potentially endogenous business model variable). These results are available upon request.
[28] The literature has expressed criticism against the stepwise regression approach when it is used as a primary method for selecting variables (e.g. Judd and McClelland 2008). Our use here is intended purely as a robustness check.

Turning to the business model indicator variable, a consistently statistically significant coefficient is estimated for the securities holding banks and the traditional commercial banks. First, securities holding banks seem to be characterised by higher credit risk in their loan portfolio compared to the rest of the banks, both with respect to realised credit risk and the default rate. This would seem to be consistent with an interpretation based on the specialisation of these banks, namely that because loan granting represents a relatively less significant component of their assets, as is seen in Figure 2, they are less well placed to monitor effectively credit risk. This result is also consistent with the finding regarding the elevated capital levels held by this cluster of banks and the need to hold capital buffers for potential losses.

Second, the traditional commercial banks are found to be exposed less to credit risk compared to the other banks. These banks exemplify the standard intermediation function, as in the seminal contribution of Ho and Sanders (1981), and can be seen to stand in contrast to the securities holding banks, which deviate significantly from the traditional form of banking intermediation. Therefore, it is intuitive that these traditional 'boring' institutions exhibit relatively low levels of loan portfolio credit risk, in contrast to banks for which loan granting is less pronounced compared to market activities.

It is also interesting that size variables, like total assets and market share are not found statistically significant with the same consistency as the business model indicators for securities holding and traditional commercial banks. Our results hint to the fact that business models may be statistically more robust determinants for credit risk compared to size indicators, which is supportive of the practice of distinguishing the different types of banks when analysing their credit risk as, for example, in Ghosh (2015). This practice is common already in the analyses of market practitioners (e.g. SNL 2013, KPMG 2016) and regulators e.g. in the Supervisory Review and Evaluation Process (SREP) followed in the eurozone (ECB 2016) and the Federal Reserve (Carlson et al., 2017).[29]

---

[29] We have also checked the relation of market based measures of risk with business model at the level of descriptive statistics, due to the lack of data. Specifically, when using the CDS spreads we find significantly higher values for the complex commercial banks. However, the sample of banks for which CDS are available is not enough for an econometric investigation (36 out of 365 banks).

Furthermore, it has been argued that market-based measures of credit risk, like SRISK, are useful to measure risk in states of the world where all equity is wiped out and subsequently balance sheet measures is more suitable to quantify potential credit risk losses (Homar et al. 2016). The comparison of market based and balance sheet based measures of risk would be however an interesting topic for further research.

**Table 1**

The dependent variable is realised credit risk, defined as impaired loans minus their specific provisions as a percentage of total loans, as observed in 2015. Real GDP growth is the y-o-y growth in end-2015. Unemployment change is the 6-year change in unemployment until end-2015. The rest of the bank-specific variables are as observed on end-2014 except from the growth rates which refer to yearly rates by end-2015. Log (total assets) refers to the logarithm of bank's total assets. Deposits/liabilities is the ratio of deposits to liabilities. Equity/assets is the ratio of equity to assets. Growth in assets is the yearly growth rate in total assets, from 2014Q4 to 2015Q4. Growth in income is the yearly growth rate in operating income. Market share is the bank's share in total assets in the country. The business model dummy is one if the bank belongs to the respective business model indicated by the column label. t-statistics are reported in parentheses.

| Dependent variable: Realised credit risk | | | | | | | | |
|---|---|---|---|---|---|---|---|---|
| | Wholesale funded banks (Cluster 1) | | Securities holding (Cluster 2) | | Traditional commercial (Cluster 3) | | Complex commercial (Cluster 4) | |
| | OLS | Country fixed effects | OLS | Country fixed effects | OLS | Country fixed effects | OLS | Country fixed effects |
| GDP growth | -0.482 (0.79) | | -0.569 (0.97) | | -0.580 (1.03) | | -0.320 (0.52) | |
| Unemployment change | 0.772* (1.94) | | 0.763* (2.01) | | 0.773** (2.22) | | 0.791* (2.09) | |
| Log (total assets) | 0.024 (1.66) | 0.038 (1.54) | 0.026* (2.18) | 0.036* (1.69) | 0.022 (1.61) | 0.032 (1.49) | 0.019 (1.25) | 0.030 (1.39) |
| Deposits/liabilities | 0.449** (2.31) | 0.577 (1.84)* | 0.455** (2.39) | 0.577* (1.83) | 0.480** (2.50) | 0.602* (1.84) | 0.455** (2.27) | 0.580* (1.78) |
| Equity/assets | -0.504 (0.89) | -0.343 (0.73) | -0.448 (0.78) | -0.307 (0.62) | -0.453 (0.82) | -0.306 (0.62) | -0.566 (1.02) | -0.395 (0.86) |
| Growth in assets | -0.527* (2.18) | -0.701 (2.13)** | -0.601** (2.57) | -0.765** (2.37) | -0.618** (2.27) | -0.775** (2.08) | -0.513* (2.06) | -0.695** (2.21) |
| Growth in income | -0.076* (2.04) | -0.060* (1.77) | -0.079* (2.08) | -0.062* (1.75) | -0.072* (1.88) | -0.055 (1.62) | -0.072* (2.05) | -0.054* (1.77) |
| Market share | -0.329 (1.09) | -0.340 (1.11) | -0.280 (0.94) | -0.267 (0.98) | -0.288 (0.96) | -0.270 (0.97) | -0.357 (1.09) | -0.360 (1.14) |
| Business model dummy | -0.033 (0.83) | -0.063 (1.27) | 0.041* (1.75) | 0.042** (2.04) | -0.042* (1.73) | -0.035* (1.90) | 0.046 (1.05) | 0.063 (1.09) |
| Constant | -0.800* (1.81) | -1.193 (1.55) | -0.863** (2.40) | -1.178* (1.69) | -0.749* (1.85) | -1.076 (1.53) | -0.696 (1.50) | -1.022 (1.44) |
| Observations | 136 | 136 | 136 | 136 | 136 | 136 | 136 | 136 |
| R-squared | 0.34 | 0.27 | 0.34 | 0.28 | 0.34 | 0.28 | 0.34 | 0.28 |

**Table 2**

The dependent variable is the default rate, defined as the ratio of defaulted loans to total loans, as observed in 2015. Real GDP growth is the y-o-y growth in end-2015. Unemployment change is the 6-year change in unemployment until end-2015. The rest of the bank-specific variables are as observed on end-2014 except from the growth rates which refer to yearly rates by end-2015. Log (total assets) refers to the logarithm of bank's total assets. Deposits/liabilities is the ratio of deposits to liabilities. Equity/assets is the ratio of equity to assets. Growth in assets is the yearly growth rate in total assets, from 2014Q4 to 2015Q4. Growth in income is the yearly growth rate in operating income. Market share is the bank's share in total assets in the country. The business model dummy is one if the bank belongs to the respective business model indicated by the column label. t-statistics are reported in parentheses.

| Dependent variable: Default rate | | | | | | | | |
|---|---|---|---|---|---|---|---|---|
| | Wholesale funded banks (Cluster 1) | | Securities holding (Cluster 2) | | Traditional commercial (Cluster 3) | | Complex commercial (Cluster 4) | |
| | OLS | Country fixed effects | OLS | Country fixed effects | OLS | Country fixed effects | OLS | Country fixed effects |
| GDP growth | -0.020 (0.05) | | 0.117 (0.28) | | -0.028 (0.08) | | -0.041 (0.10) | |
| Unemployment change | 0.739** (2.53) | | 0.793** (2.65) | | 0.735** (2.73) | | 0.720** (2.39) | |
| Log (total assets) | -0.003 (0.38) | 0.005 (1.19) | 0.000 (0.04) | 0.006 (1.40) | -0.003 (0.47) | 0.005 (1.11) | -0.003 (0.46) | 0.005 (1.16) |
| Deposits/liabilities | 0.058 (1.33) | 0.032 (1.01) | 0.061 (1.44) | 0.041 (1.32) | 0.063 (1.43) | 0.032 (1.02) | 0.053 (1.15) | 0.031 (0.98) |
| Equity/assets | -0.002 (0.02) | 0.026 (0.31) | -0.052 (0.55) | -0.007 (0.09) | -0.018 (0.20) | 0.023 (0.28) | -0.005 (0.05) | 0.026 (0.32) |
| Growth in assets | -0.154** (2.92) | -0.147** (2.50) | -0.172*** (3.11) | -0.145** (2.56) | -0.187*** (3.31) | -0.147** (2.51) | -0.158** (2.90) | -0.145** (2.48) |
| Growth in income | 0.017 (1.07) | 0.026** (2.02) | 0.025 (1.20) | 0.027** (2.19) | 0.020 (1.33) | 0.026** (2.04) | 0.016 (0.97) | 0.026** (2.02) |
| Market share | 0.169 (1.46) | -0.044 (0.57) | 014 0.147 (1.28) | -0.048 (0.64) | 0.163 (1.52) | -0.043 (0.55) | 0.164 (1.45) | -0.045 (0.59) |
| Business model dummy | 0.000 (0.02) | -0.005 (0.39) | 0.049* (1.73) | 0.027** (2.13) | -0.036** (2.16) | -0.015* (1.73) | 0.014 (0.76) | -0.000 (0.02) |
| Constant | 0.087 (0.53) | -0.031 (0.26) | 0.021 (0.14) | -0.055 (0.48) | 0.108 (0.74) | -0.020 (0.18) | 0.097 (0.60) | -0.026 (0.22) |
| Observations | 121 | 121 | 121 | 121 | 121 | 121 | 121 | 121 |
| R-squared | 0.48 | | 0.52 | | 0.51 | | 0.48 | |

**Table 3**
The dependent variable is realised credit risk, defined as impaired loans minus their specific provisions as a percentage of total loans, as observed in 2015. Real GDP growth is the y-o-y growth in end-2015. Unemployment change is the 6-year change in unemployment until end-2015. The rest of the bank-specific variables are as observed on end-2014 except from the growth rates which refer to yearly rates by end-2015. Log (total assets) refers to the logarithm of bank's total assets. Deposits/liabilities is the ratio of deposits to liabilities. Equity/assets is the ratio of equity to assets. Growth in assets is the yearly growth rate in total assets, from 2014Q4 to 2015Q4. Growth in income is the yearly growth rate in operating income. Market share is the bank's share in total assets in the country. The business model dummy is one if the bank belongs to the respective business model indicated by the column label. t-statistics are reported in parentheses. For each business model regression, the first column presents results with an instrumental variables estimation while the second column with a backward stepwise approach.

Dependent variable: Realised credit risk
Robustness checks

|  | Wholesale funded banks (Cluster 1) | | Securities holding (Cluster 2) | | Traditional commercial (Cluster 3) | | Complex commercial (Cluster 4) | |
| --- | --- | --- | --- | --- | --- | --- | --- | --- |
|  | IV GMM | Stepwise | IV GMM | Stepwise | IV GMM | Stepwise | IV GMM | Stepwise |
| GDP growth | -0.716 (0.57) |  | -0.937 (1.12) | -1.075* (2.18) | -1.230 (1.08) | -1.090** (2.24) | 0.343 (0.29) |  |
| Unemployment change | 0.664 (1.30) | 0.931** (2.95) | 0.660* (1.87) | 0.598* (1.85) | 0.637* (1.75) | 0.609* (2.02) | 0.631 (1.35) | 0.931** (2.95) |
| Log (total assets) | 0.034 (0.90) |  | 0.036** (2.25) | 0.029** (2.78) | 0.023 (1.57) | 0.023** (2.22) | 0.005 (0.19) |  |
| Deposits/liabilities | 0.432*** (2.70) | 0.285* (1.88) | 0.455** (2.55) | 0.467** (2.26) | 0.569*** (3.05) | 0.496** (2.37) | 0.458* (1.82) | 0.285* (1.88) |
| Equity/assets | -0.496 (1.13) | -0.876 (1.58) | -0.321 (0.61) |  | -0.261 (0.49) |  | -1.068 (1.31) | -0.876 (1.58) |
| Growth in assets | -0.530* (1.83) | -0.325 (1.37) | -0.764*** (2.70) | -0.633** (2.40) | -0.949*** (3.23) | -0.647* (2.17) | -0.406 (1.14) | -0.325 (1.37) |
| Growth in income | -0.078* (1.88) | -0.064 (1.76) | -0.086 (1.97)** | -0.072* (2.02) | -0.062 (1.38) | -0.064* (1.81) | -0.049 (0.94) | -0.064 (1.76) |
| Market share | -0.411 (0.85) |  | -0.236 (0.81) |  | -0.243 (0.71) |  | -0.831 (1.51) |  |
| Business model dummy | -0.127 (0.38) |  | 0.130* (1.71) | 0.052** (2.74) | -0.192** (2.05) | -0.050** (2.22) | 0.437 (1.32) |  |
| Constant | -0.974 (1.20) | -0.114 (0.99) | -1.131** (2.30) | -0.987** (2.61) | -0.787* (1.73) | -0.841** (2.21) | -0.333 (0.45) | -0.114 (0.99) |
| Observations | 136 | 136 | 136 | 136 | 136 | 136 | 136 | 136 |
| R-squared | 0.32 | 0.32 | 0.31 | 0.32 | 0.24 | 0.32 | 0.26 | 0.32 |

**Table 4**
The dependent variable is default rate, defined as the ratio of defaulted loans to total loans, as observed in 2015. Real GDP growth is the y-o-y growth in end-2015. Unemployment change is the 6-year change in unemployment until end-2015. The rest of the bank-specific variables are as observed on end-2014 except from the growth rates which refer to yearly rates by end-2015. Log (total assets) refers to the logarithm of bank's total assets. Deposits/liabilities is the ratio of deposits to liabilities. Equity/assets is the ratio of equity to assets. Growth in assets is the yearly growth rate in total assets, from 2014Q4 to 2015Q4. Growth in income is the yearly growth rate in operating income. Market share is the bank's share in total assets in the country. The business model dummy is one if the bank belongs to the respective business model indicated by the column label. t-statistics are reported in parentheses. For each business model regression, the first column presents results with an instrumental variables estimation while the second column with a backward stepwise approach.

| Dependent variable: Default rate Robustness checks | | | | | | | | |
|---|---|---|---|---|---|---|---|---|
| | Wholesale funded banks (Cluster 1) | | Securities holding (Cluster 2) | | Traditional commercial (Cluster 3) | | Complex commercial (Cluster 4) | |
| | IV GMM | Stepwise | IV GMM | Stepwise | IV GMM | Stepwise | IV GMM | Stepwise |
| GDP growth | -0.044 (0.13) | | 0.057 (0.18) | | -0.054 (0.10) | | -0.359 (0.55) | |
| Unemployment change | 0.731*** (4.85) | 0.741** (2.76) | 0.769*** (4.81) | 0.784** (2.88) | 0.722*** (5.58) | 0.738*** (2.99) | 0.421* (1.80) | 0.741** (2.76) |
| Log (total assets) | -0.002 (0.33) | | -0.001 (0.28) | | -0.004 (0.87) | | -0.011 (1.47) | |
| Deposits/liabilities | 0.056* (1.74) | 0.071 (1.46) | 0.060** (2.15) | 0.061 (1.50) | 0.080* (1.70) | 0.077 (1.65) | -0.019 (0.34) | 0.071 (1.46) |
| Equity/assets | -0.008 (0.09) | | -0.030 (0.32) | | -0.074 (0.69) | | -0.050 (0.31) | |
| Growth in assets | -0.157** (2.41) | -0.109* (1.99) | -0.164** (2.39) | -0.118* (1.93) | -0.301*** (3.09) | -0.135* (2.11) | -0.218** (2.07) | -0.109* (1.99) |
| Growth in income | 0.018 (0.98) | | 0.021 (1.02) | | 0.031 (1.24) | | -0.011 (0.37) | |
| Market share | 0.161 (1.75)* | 0.132 (1.58) | 0.156** (2.04) | 0.145 (1.68) | 0.144 (1.43) | 0.123 (1.43) | 0.097 (0.61) | 0.132 (1.58) |
| Business model dummy | -0.015 (0.17) | | 0.028* (1.77) | 0.043* (1.80) | -0.160*** (3.91) | -0.035** (2.19) | 0.232*** (2.90) | |
| Constant | 0.074 (0.65) | -0.005 (0.15) | 0.050 (0.51) | -0.008 (0.21) | 0.181 (1.30) | 0.002 (0.05) | 0.257 (1.33) | -0.005 (0.15) |
| Observations | 121 | 121 | 121 | 121 | 121 | 121 | 121 | 121 |
| R-squared | 0.48 | 0.47 | 0.51 | 0.51 | 0.51 | 0.51 | 0.49 | 0.47 |

## 6. Conclusion

We present the first study that makes use of an exceptionally granular data set on European banks in order to infer the types of existing business models. We have adopted a data driven clustering approach which combines optimally the classification of banks with data reduction, enhanced with an 'outlier' banks detection component in order to avoid the 'contamination' of the derived clusters with very specialised institutions. Our approach minimises the impact of the researcher's priors on the results while taking into account the full range of banks' activities and avoiding misclassifications due to the use of a narrow set of balance sheet ratios.

The results provide an anatomy of the Euro area banking sector and indicate the co- existence of distinct business models. We label the four business models identified by our method as "traditional commercial", "complex commercial", "wholesale funded" and "securities holding" banks. Specialised institutions such as state owned entities aimed at refinancing loans to semi-public and public entities, central clearing counterparties or banks in a run-down mode, are identified as outliers by the clustering algorithm.

The statistical analysis identifies two main factors as the most efficient composite variables to discriminate banks: a level factor representing the presence of "standard" asset and liability items, with the notable exception of trading assets, and a contrast factor which represents the imbalance in the presence of loans on the asset side compared to "standard liabilities" with the latter including deposits and issued debt. These factors represent banks' activities and are robust to the presence of the various forms through which these activities are reflected in the bank's balance sheet e.g. whether a bank provides credit to non-financial corporations via regular loans or credit lines.

Moreover, our econometric investigation provides empirical evidence of a relationship between credit risk in the loan portfolio and the choice of the business model. We find that traditional commercial banks present significantly lower credit risk compared to the other business models. In addition, securities holding banks are characterized by higher levels of credit risk. These results prove to be largely robust to alternative specifications and when controlling for the endogeneity of the business model.

Our results reflect the history of macroeconomic developments and policy decisions which have taken place during the crisis period while dynamic effects may also be important e.g. like migration of banks towards different business models, and should be further investigated. For example, both the ECB Financial Stability Review (ECB 2015) and the IMF Global Financial Stability Report (IMF 2015) point to structural business model changes in the aftermath of the global financial crisis, in a context characterised by low inflation, weak profitability and large, albeit heterogeneously distributed, non-performing loans.

These results are highly relevant for monetary policy, micro-prudential banking supervision and the design of macro-prudential policy. Specifically, asymmetries in the monetary policy transmission mechanism across countries, as identified for example by Barigozzi et al. (2014), could be also partly explained by the prevalence of different banks' business models in single countries. Empirical research points to the significance that should be attached to the heterogeneity of banks' business models when designing micro-prudential supervision (Blundell-Wignall et al., 2014) and this is already reflected in the supervision process in many jurisdictions (e.g. see ECB, 2016). Furthermore, the paper has provided evidence that business models represent determinants of credit risk, which is especially relevant for banking supervision. It would be of great interest to investigate a possible link between business models and measures of systemic risk that would inform also the calibration of structural macro-prudential capital tools e.g. the systemic risk buffer (SRB) or the other systemically important institutions (O-SII) buffer. Finally, it would be of great interest to investigate whether the business models aspect could shed light on the relationship between competition and risk-taking in banking (Boyd and De Nicolò, 2005) e.g. by influencing the correlation structure of losses (as for example Hakenes and Schnabel, 2011).

# Appendix

## A – FINREP templates used as input set

Table A-1 presents the templates which are used as the input set along with the number of variables from each template. The EBA templates along with the definitions of the contained data can be found at the EBA website: https://www.eba.europa.eu/regulation-and-policy/supervisory-reporting.

**Table A-1:** Contents of the templates defining the input set

| Template code (as defined by the EBA) | Contents |
|---|---|
| F 01.01 | Assets |
| F 04.01 | Assets held for trading |
| F 04.02 | Assets designated at fair value through profit or loss |
| F 04.03 | Available-for-sale assets (carrying amount) |
| F 04.04 | Loans and receivables & held-to-maturity assets |
| F 05.00 | Loans and advances by product (on demand, credit card, leases, loans etc) |
| F 08.01.a | Liabilities |
| F 09.01 | Off-balance items (loan commitments and guarantees) |
| F 10.00 | Derivatives – trading |
| F 11.01 | Derivatives – hedge accounting |

F 01.01 and all F 04 templates provide the breakdown of assets across accounting portfolios, with additional breakdowns on instruments and counterparties. F 05 provides the breakdown of loans by product (credit card loans, collateralized loans, project finance etc). The liability side is covered by the F 08 template, which breaks down liabilities by accounting portfolio (the largest percentage of banks' liabilities are valued at amortised cost), instrument and counterparty. Off-balance sheet items, primarily loan commitments and guarantees are contained in template F 09.01. Finally, templates F 10 and F 11.01 provide detailed information on derivatives, distinguishing between trading and hedge accounting derivatives. The information is further broken down by type of derivatives (interest rate, equity, foreign exchange, credit and commodity) and by the type of market in which the derivatives are traded (OTC or organised markets).

**B – Input set selection**

The initial set of variables contains a number of highly correlated variables. We would like to automatically select the ones which are more "fundamental" in the sense of being more relevant for supervisory purposes; for example, when there is a variable like "notional amount of total derivatives" and one of its subcategories like "notional amount of OTC derivatives", we would prefer to keep the broader category on the condition that it is more related to the set of the remaining variables. This selection of variables is also subject to the condition that we would like to exclude pairs of variables with the absolute level of correlations above a threshold, which was set to 0.95, in order to avoid bias in the results.

We define a measure of the "importance" of each variable within the data set in order to operationalise the above selection criteria. The "importance" $I(j)$ of each variable $j$ is defined as the linear combination of the correlation absolute values with the other variables of the input set:

$$I(j) = \sum_{\substack{k=1,\ldots,P, \\ k \neq j}} |Corr(j,k)|$$

Consequently, we order the variables in a non-increasing order based on their $I(j)$. Whenever the predefined level of correlation $C^*$ is exceeded, then only one of the two correlated variables is retained, specifically the one with a higher level of $I(j)$.

This step represents a data preparation step motivated by the nature of our initial dataset and is not related to the dimensionality reduction component of the clustering algorithm that we employ to derive the banks' classification into business models.

## C – Clustering algorithm

The clustering algorithm which minimises Eq. (4) belongs to the class of Alternated Least Squares (ALS) algorithms (Vichi and Kiers, 2001).

In Step 1, we minimize $F(\boldsymbol{A}, \boldsymbol{U}, \overline{\boldsymbol{Y}})$ with respect to $\boldsymbol{U}$ given the values of $\boldsymbol{A}$ and $\overline{\boldsymbol{Y}}$. For each row $i$ of $U$ we set $U_{ij} = 1$, if $F(\boldsymbol{A}, U_{ij}) = \min\{F(\boldsymbol{A}, [U_{iv}]): v = 1, \dots, m\}$ and $U_{ij} = 0$ otherwise.

In Step 2, $F(\boldsymbol{A}, \boldsymbol{U}, \overline{\boldsymbol{Y}})$ is minimized keeping fixed $\boldsymbol{U}$, to update *jointly* $\boldsymbol{A}$ and $\overline{\boldsymbol{Y}}$. Among all the linear combinations of $\boldsymbol{X}$, the ones closer to the centroids (in the transformed space) are derived by taking the first $m$ eigenvectors of $\boldsymbol{X}'(\boldsymbol{U}\,(\boldsymbol{U}'\boldsymbol{U})^{-1}\boldsymbol{U}' - \boldsymbol{I}_n)\boldsymbol{X}$ (see Ten Berge 1993). From the optimal $\boldsymbol{A}$, we then update $\overline{\boldsymbol{Y}}$ from the expression $\overline{\boldsymbol{Y}} = (\boldsymbol{U}'\boldsymbol{U})^{-1}\boldsymbol{U}'\boldsymbol{X}\boldsymbol{A}$.

In Step 3, we compute $F(\boldsymbol{A}, \boldsymbol{U}, \overline{\boldsymbol{Y}})$ for the current values of $\boldsymbol{U}$, $\boldsymbol{A}$ and $\overline{\boldsymbol{Y}}$. If $F$ has decreased, we go again with Step 1 and 2. Otherwise, the process has converged. In the latter case, we keep the values of $\boldsymbol{U}$, $\boldsymbol{A}$ and $\overline{\boldsymbol{Y}}$ from the previous iteration.

The algorithm as described above starts with some given values of $\boldsymbol{A}$ and $\overline{\boldsymbol{Y}}$. In order to avoid being trapped by a local optimum, we have to cover the parameter space. For this reason, a procedure for choosing a number of random initial points in a rational way is applied.

Matrix $\boldsymbol{A}$ is initialized in the following way. First, $K=100$ random permutation matrices $\boldsymbol{P}_k$, $k = 1, \dots, K$, are generated. The Gram-Schmidt algorithm is applied on each of these matrices, in order to make them orthogonal. Our input matrix is the $382 \times 382$ unbiased sample covariance matrix $\boldsymbol{C} = \frac{1}{n-1}\boldsymbol{X}'\boldsymbol{X}$. Its spectral decomposition is $\boldsymbol{V}_r\,\boldsymbol{D}_r\boldsymbol{V}_r{'}$, where $\boldsymbol{V}_r$ is the matrix of eigenvectors and $\boldsymbol{D}_r$ is the matrix of eigenvalues. Thus, we post-multiply the random permutation matrix $\boldsymbol{P}_k$ by $\boldsymbol{V}_r$. We run this initial procedure for $K=100$ times obtaining the same number of initial estimates $\boldsymbol{A}_{0,k} = \boldsymbol{P}_k\boldsymbol{V}_r$, $k = 1, \dots, K$.

For each of the 100 random $\boldsymbol{A}_{0,k}$, the group matrix $\boldsymbol{U}$ is initialised by computing for each bank the relevant quantity $t_i = n\,diff_i(\boldsymbol{C}_F)^{-1}diff_i'$, $i = 1, \dots, 365$, where $\boldsymbol{C}_F$ is the computed covariance matrix on the columns of $\mathbf{F}_{0,k} = \boldsymbol{X}\boldsymbol{A}_{0,k}$, and we set $diff_i = \mathrm{F}_{0,k,i} - \overline{\mathrm{F}}_{0,k}$ for $k = 1, \dots, 100$ (denoting by $\mathrm{F}_{0,k,i}$ the i-th row of the matrix $\mathrm{F}_{0,k}$, and by $\overline{\mathrm{F}}_{0,k}$ the mean across rows). Then, given the number of groups $c$, the $2c$ quantiles of the distribution of $t_i$ across banks are computed. The distance between each score in the low-dimensional space ($\mathrm{F}_i$) and the first, third, fifth and seventh (i.e. the odd quantiles, $1, 3, \dots 2c - 1$ in general) quantile is computed. So, each bank is assigned to the closer quantile, thus originating c initial groups for each $A$. The cluster corresponding to the seventh quantile is smaller than the other three, which are instead similar.[30] Finally, at each run $r$, the centroids $\bar{Y}$ are updated as $(U'U)^{-1}U'XA$.

An outlier detection component is incorporated in the above algorithm. Let us suppose we have estimated the reduced space $\widehat{F} = X\widehat{A}$, which is a n x r matrix. We can exploit these low-dimensional representations of the objects for outlier detection purposes. If we call $\hat{F}_i$ the i-th row of the matrix $\widehat{F}$, and $\bar{F}_i$ the mean of F across the cluster to which the i-th observation belongs, we can define $\text{diff}_i = F_i - \bar{F}_i$ for each $i = 1, \ldots, 365$. We compute $C_F$, the unbiased sample covariance matrix for these low-dimensional data. The quantity $t_i = n \, \text{diff}_i (C_F)^{-1} \text{diff}_i'$ is the relevant one to define the outliers. Under the normality assumption for $X$, we would have $t \sim T_{p,n-1}^2$ which is called Hotelling's T (1931). In this case, since this assumption is violated, we can compute these values for all observations, and then derive, for instance, as a threshold the $100 \times (1-\alpha)\%$ -th percentile of the empirical distribution of $t_i$ across the sample. This value is our empirical quantile. The observations having a value for $t_i$ exceeding it are flagged as outliers.

This calculation is included directly in the iterative clustering procedure. At each step, if an observation $o_i$ is identified as an outlier, we set $U_{ij} = 0$ for j= 1, ..., c, such that the observations pervasively distant from the centroid of any cluster are identified. In this way, outliers are excluded from the computation of the coefficient matrix $A$ and the centroid matrix $\bar{Y}$.

The number of clusters and the number of factors have to be determined simultaneously in our approach. For this selection we used a step-wise approach, utilising a partition-based criterion to select the number of clusters combined with search in the number of factors space and then selecting the number of factors. It has to be noted that in the current setting which involves simultaneous clustering and dimensionality reduction, the approaches of deciding on the number of factors based on the initial data set and its covariance matrix are not suitable. In contrast, our adopted approach involves first selecting the number of clusters as recommended in Vichi and Kiers 2001.

A measure of the fitness of a specific clustering is given by the sum of within-cluster distances to centroids:

---

[30] The means of the percentages of cluster belongings across our 100 runs are, respectively 0.260, 0.254, 0.269, and 0.217. We can see that the last group tends to be penalised, while the third is the largest by a small margin. This is consistent with the empirical distribution of $t_i$ across banks.

$$W_c(P, \overline{Y}) = \sum_{j=1}^{c} \sum_{i \in P_j} d(i, \bar{y}_j)$$

where $P=\{P_1, P_2,…, P_c\}$ is a partition of c clusters and $d$ is the Euclidean distance measure. Hartigan (Hartigan 1975) has proposed a heuristic rule to select the number of clusters which has been shown to be effective in subsequent simulation studies (see e.g. Chiang and Mirkin 2010). The idea of the method is that when the optimal c* is used a decrease of W with respect to c<c* will be observed because "coarse" clusters defined by c<c* will be split further while when c>c* the value of W will be relatively less volatile given that "optimal" clusters will be split in a random way. Hartigan suggested the calculation of the metric

$$H_c = (W_c/W_{c+1} - 1)(p - c - 1)$$

and selecting *c* when a large increase in this metric is observed when *c+1* are used.[31]

We provide two MATLAB functions performing the trimmed factorial k-means procedure, 'tfkm.m' and 'tfkm_alpha.m'. The first one performs the procedure selecting the rank and the number of clusters by maximizing Hartigan's statistics. The second one also selects the trimming proportion via the same method. Details can be found at the link https://github.com/MatFar88/A-clustering-methodology-for-European-banks-business-models.

---

[31] Hartigan suggested specifically that when $H_c > 10$, the number of clusters should be selected to be equal to *c+1*. Chang and Mirkin (2010) find that this criterion works well also for different values than 10.

## D – Sample characteristics per country

Our sample is quite representative of the eurozone countries. We compare the amounts represented by the banks in our sample, with the ECB Consolidated Banking Statistics. Due to the existence of subsidiaries and the consolidation at the entity level of all our data, their aggregation leads to double counting at the national level, therefore we get figures which exceed 100% for some countries.

Table D-1: Number of banks per country and total assets (in billion euros) included.

| Country | Number of banks | Assets of banks included | Percentage of consolidated assets |
|---|---|---|---|
| Austria | 22 | 993.3 | 92% |
| Belgium | 25 | 1549.6 | 141% |
| Cyprus | 4 | 48.0 | 63% |
| Germany | 24 | 4698.2 | 67% |
| Estonia | 6 | 15.5 | 70% |
| Spain | 53 | 4361.8 | 122% |
| Finland | 15 | 502.2 | 88% |
| France | 48 | 8396.1 | 117% |
| Greece | 5 | 353.0 | 96% |
| Ireland | 7 | 359.5 | 71% |
| Italy | 61 | 2697.2 | 100% |
| Lithuania | 6 | 65.9 | 273% |
| Luxembourg | 23 | 434.7 | 54% |
| Latvia | 13 | 25.6 | 83% |
| Malta | 3 | 15.3 | 29% |
| Netherlands | 18 | 2268.3 | 90% |
| Portugal | 16 | 377.4 | 89% |
| Slovenia | 8 | 27.3 | 66% |
| Slovakia | 8 | 48.9 | 78% |

**E – Description of country-level variables and robustness checks**

Macroeconomic variables. Unemployment rate and the annual real GDP growth were taken from the Eurostat. Besides the unemployment rate for end-2014 and the real GDP growth rates for 2014, the average unemployment, the (annualised) growth of unemployment from 2009 to 2016 was computed, as a measure of the macroeconomic stress of each country during the crisis. In addition, the average real growth for the 5 years before 2014 is also calculated as a measure of past macroeconomic conditions.

Size of national financial sectors (total assets). We use Balance Sheet Items (BSI data) from the ECB which are based on the consolidated balance sheet of Monetary Financial Institutions (MFIs) – including banks, other deposit-taking firms and money market funds – but excluding central banks. When the value of assets of each bank in our sample is divided by this amount, a measure for the systemic importance of the bank within each national financial sector is provided.

Bank credit-to-market ratio. We use World Bank Data from the financial development and structure data set as described in Beck *et al.* (2000) and Čihák *et al.* (2012). Namely, we use data on market capitalization of domestic firms (excluding investment funds and other companies which only hold shares of other listed companies, to avoid double-counting) and on domestic credit to the private sector provided by banks (which includes both loans and securities), in both cases for end-2014. Bank credit is reported on a host–country basis i.e. includes also subsidiaries and branches of foreign banks. Banks are defined with respect to their license to receive retail deposits. For Estonia, Finland, Lithuania, Latvia and Slovakia, the market capitalization ratio was not available, therefore data on the stock market capitalization to GDP were used (from the same World Bank database). As a robustness check and to account for the structural features of each economy, rather the snapshot at end-2014, the average figures for the last six years starting from 2008, for both market capitalization and bank credit, were also calculated and used as an alternative measure.

Financial structure indicators. Concentration of the national banking sectors is taken from the ECB Statistical Data Warehouse (SDW). Two alternative measures are used, namely the Herfindahl index for the total assets of credit institutions and the share of total assets of the five (5) largest credit institutions at the end of 2014.

# References

Albertazzi, U., Gambacorta, L., 2009. Bank profitability and the business cycle. Journal of Financial Stability 5, 393-409.

Altunbas, Y., Binici, M., Gambacorta, L., 2018. Macroprudential policy and bank risk. Journal of International Money and Finance 81, 203-220.

Amel, D. F., Rhoades, S. A., 1988. Strategic groups in Banking. Review of Economics and Statistics 70, no. 4, 685-689.

Anastasiou, D., Louri, H., Tsionas, M., 2016. Determinants of non-performing loans: Evidence from Euro-area countries. Finance Research Letters 18, 116-119.

Angrist, J., Pischke, J., 2009. Mostly harmless econometrics: An empiricist's companion. New Jersey: Princeton University Press.

Ayadi, R., De Groen, W., 2014. Banking business models monitor 2014. Centre for European Policy Studies and International Observatory on Financial Services Cooperatives.

Ayadi, R., De Groen W., Sassi, I., Mathlouthi, W., Rey, H., Aubry, O., 2015. Banking business models monitor 2015. Alphonse and Dorimène Desjardins International Institute for Coopeatives & International Research Centre on Cooperative Finance.

Barigozzi, M., Conti, A., Luciani, M. 2014. Do euro area countries respond asymmetrically to the common monetary policy? Oxford Bulletin of Economics and Statistics 76, no. 5, 693-712.

Becchetti, L., Ciciretti, R., Paolantonio, A., 2016. The cooperative bank difference before and after the global financial crisis. Journal of International Money and Finance 69, 224-246.

Beltratti, A., Stulz, R.M., 2012. The crecit crisis around the globe: Why did some banks perform better? Journal of Financial Economics 105, 1-17.

Beck, T., Demirgüç-Kunt A. and Levine R., 2000. A new database on financial development and structure. World Bank Economic Review. 14, 597-605.

Berger, A. N., DeYoung, R., 1997. Problem loans and cost efficiency in commercial banks. Journal of Banking and Finance 21, 849-870.

Berger, A. N., Mester, L. J., 2003. Explaining the dramatic changes in performance of US banks: technological change, deregulation, and dynamic changes in competition. Journal of Financial Intermediation 12(1), 57-95.

Berger, A. N., Bouwman, C. H. S., 2013. How does capital affect bank performance during financial crises? Journal of Financial Economics 109, 146-176.

Blundell-Wignall, A., Atkinson, P., Roulet, C., 2014. Bank business models and the Basel system: Complexity and interconnectedness. OECD Journal: Financial Market Trends 2014, vol. 2013/2.

Boyd, J., Gertler, M., 1994. The role of large banks in the recent US banking crisis. Federal Reserve Bank of Minneapolis Quarterly Review 18, 1-21.

Calinski, T., Harabasz, J., 1974. A dendrite method for cluster analysis, Communications in Statistics 3, 1-27

Carlson, M., Shatto, M., Warusawitharana, M., 2017. Matching banks by business model, geography and size: A dataset. FEDS Notes, August 8 2017.

Carney, M., 2015. Breaking the tragedy of the horizon – climate change and financial stability. Speech at Lloyd's, London, 29 September.

Caves, R. E., Porter, M. E., 1977. From entry barriers to mobility barriers: Conjectural decisions and contrived deterrence to new competition. Quarterly Journal of Economics 91, no. 2, 241-262.

Chiang, M.M., Mirkin, B., 2010. "Intelligent choice of the number of clusters in K-Means clustering: An experimental study with different cluster spreads". Journal of Classification. 27, 3-40.

Cuesta-Albertos, J.A., Gordaliza, A., Matrán C., 1997. Trimmed k-means: An attempt to robustify quantizers. Annals of Statistics 25, 553-576.

Demirgüç-Kunt, A., and Huizinga, H., 1999. Determinants of commercial bank interest margins and profitability: Some international evidence. The World Bank Economic Review 13, 379-408.

DeSarbo, W.S., Jedidi, K., Cool, K., Schendel, D., 1990. Simultaneous multidimensional unfolding and cluster analysis: an investigation of strategic groups. Marketing Letters 2, 129-146.

De Soete G. and Caroll J. D, 1994. K-means clustering in a low-dimensional Euclidean space, in E. Diday, Y. Lechevallier, M. Schader, P. Bertrand and B. Burtschy (Eds.). New approaches in classification and data analysis. Berlin, Springer-Verlag.

Draghi, M., 2016. Introductory statement at the European Parliament's Economic and Monetary Affairs Committee. Brussels, 15 February.

Eichengreen, B., Gupta, P., 2013. The financial crisis and Indian banks: Survival of the fittest? Journal of International Money and Finance 39, 138-152.

Elsas, R., Hackethal, A., Holzhäuser, M., 2010. The anatomy of bank diversification. Journal of Banking and Finance 34, 1274-1287.

Ennis, H., Malek, H., 2005. Bank risk of failure and the too-big-to-fail policy. Federal Reserve Bank of Richmond Economic Quarterly 91(2), 21-44.

European Central Bank, 2015. Financial Stability Review. November 2015. European Central

Bank, 2016. SSM SREP: Methodology booklet. 2016 edition.

Freixas, X., Rochet, J., 2013. Taming systemically important financial institutions. Journal of Money, Credit and Banking 45, 37-58.

Ghosh, A., 2015. Banking-industry specific and regional economic determinants of non-performing loans: Evidence from US states. Journal of Financial Stability 20, 93-2014.

Hakenes, H., and Schnabel, I., 2011. Capital regulation, bank competition, and financial stability. Economic Letters 113, no. 3, 256-258.

Halaj, G., Zochowski, D., 2009. Strategic groups and bank's performance. Financial Theory and Practice 33, 153-186.

Hartigan, J.A., 1975. Clustering Algorithms. New York: J. Wiley & Sons.

Hastie, T., Tibshirani, R., Friedman, J., 2009. The Elements of Statistical Learning: Data mining, Inference, and Prediction. New York: Springer.

Ho, T., Saunders A., 1981. The determinants of bank interest margins: Theory and empirical evidence. The Journal of Financial and Quantitative Analysis 16, 581-600.

Homar, T., Kick, H., Salleo, C., 2016. Making sense of the EU wide stress test: A comparison with the SRISK approach. ECB Working Paper Series No 1920.

Hotelling, H., 1931. The generalization of Student's ratio. Annals of Mathematical Statistics 2 (3), 360-378.

Hotelling, H., 1933. Analysis of a Complex of Statistical Variables Into Principal Components. Journal of Educational Psychology 24, 417-441 and 498-520.

Hryckiewicz, A., Kozlowski, L. 2017. Banking business models and the nature of financial crisis. Journal of International Money and Finance 71, 1-24.

Hunt, M. S., 1972. Competition in the major home appliance industry, 1960-1970. Unpublished PhD dissertation, Business Economics Committee, Harvard University.

IMF 2015. Global Financial Stability Report. April 2015.

Judd, C., McClelland, G., 2008. Data analysis: A model comparison approach. Routledge, New York.

Klein, N., 2013. Non-performing loans in CESEE: Determinants and impact on macroeconomic performance. IMF Working Paper Series 13/72.

Köhler, M. 2015. Which banks are more risky? The impact of business models on bank stability. Journal of Financial Stability 16, 195-212.

KPMG 2016. Peer analysis: Predicting supervisory challenges. May 2016.

Laeven L., Levine, R., 2007. Is there a diversification discount in financial conglomerates? Journal of Financial Economics 85, 331-367.

Lin, T.-I., McLachlan, G. J., and Lee, S. X., 2016. Extending mixtures of factor models using the restricted multivariate skew-normal distribution. Journal of Multivariate Analysis 143, 398-413.

Louzis, D. P., Vouldis, A. T., and Metaxas, V. L., 2012. Macroeconomic and bank-specific determinants of non-performing loans in Greece: A comparative study of mortgage, business and consumer loan portfolios. Journal of Banking and Finance 36, 1012-1027.

Lucas, A., Schaumburg, J., Schwaab, B. 2017. Bank business models at zero interest rates. Journal of Business and Economic Statistics (forthcoming)

MacQueen, J., 1967. Some methods for classification and analysis of multivariate observations Proc. Fifth Berkeley Symp. on Math. Statist. and Prob. Vol. 1 (Univ. of Calif. Press, 1967), 281-297.

McLachlan, G.J.; Peel, D. (2000). Finite Mixture Models. Wiley.

Mergaerts, F., Vennet, R. D., 2016. Business models and bank performance: A long-term perspective. Journal of Financial Stability 22, 57-75.

Murray, P. M., Browne, R. P, and McNicholas, P. D., 2014a. Mixtures of skew-t factor analyzers. Computational Statistics & Data Analysis 77, 326-335.

Murray, P. M., McNicholas, P. D. and Browne, R. P., 2014b. A mixture of common skew-t factor analysers. Stat 3, 68-82.

Nkusu, M., 2011. Nonperforming loans and macrofinancial vulnerabilities in advanced economies. IMF Working Paper 11/161.

Porter, M. E., 1979. The structure within industries and companies' performance. Review of Economics and Statistics, 61, 214-227.

Quagliariello, M., 2007. Banks' riskiness over the business cycle: A panel analysis on Italian intermediaries. Applied Financial Economics 17, 119-138.

Reger, R. K., Huff, A. S. (1993). Strategic groups: A cognitive perspective. Strategic Management Journal, 14, 103-123.

Roengpitya, R, Tarashev, N, and Tsatsaronis, K, 2014. Bank business models. BIS Quarterly Review. December 2014, 59-65.

Rousseeuw, P. J., Leroy, A., 2003. Robust Regression and Outlier Detection. John Wiley, New Jersey.

Rousseeuw, P. J., Van Driessen, K., 2000. An algorithm for positive-breakdown regression based on concentration steps. Data Analysis, 335-346.

SNL, 2013. Fundamentals of Peer Analysis, January 2013.

ten Berge, J.M., 1993. Least squares optimization in multivariate analysis. Leiden, The Netherlands. DSWO Press, Leiden University.

Terada, Y., 2015. Strong consistency of factorial k-means clustering. Annals of the Institute of Statistical Mathematics, 67(2), 335–357.

Tywoniak, S., Galvin, P., Davies, J., 2007. New Institutional Economics' Contribution to Strategic Groups Analysis. Managerial and Decision Economics 28, 213-228.

Vichi, M., Kiers, H.A.L., 2001. Factorial k-means analysis for two-way data. Computational Statistics and Data analysis 37, 49-64.

Ward, J.H. 1963. Hierarchical Grouping to optimize an objective function. Journal of the American Statistical Association 58, 236-244.

Wheelock, D.C., Wilson, P.W., 2012. Do large banks have lower costs? New estimates of returns to scale for U.S. banks. Journal of Money, Credit and Banking 44, 171-199.

Yang, L., Sijia X., and Weixin Y., 2017. Robust fitting of mixtures of factor analyzers using the trimmed likelihood estimator. Communications in Statistics-Simulation and Computation 46(2), 1280-1291.

Yellen, J. (2012). Testimony before the Committee on Financial Services, U.S. House of Representatives, Washington, 4 November.